\documentclass[lettersize,journal]{IEEEtran}
\usepackage{amsmath,amsfonts}
\usepackage{algorithm} 
\usepackage{algorithmic}  
\usepackage[algo2e]{algorithm2e}

\usepackage{array}
\usepackage[caption=false,font=normalsize,labelfont=sf,textfont=sf]{subfig}
\usepackage{textcomp}
\usepackage{stfloats}
\usepackage{url}
\usepackage{verbatim}
\usepackage{graphicx}
\usepackage{cite}
\usepackage[usestackEOL]{stackengine}
\usepackage{amssymb}
\usepackage{svg}

\usepackage{xcolor}

\usepackage{color,soul}         

\usepackage{tablefootnote}      
\begin{document}

\title{Unrolled Architectures for High-Throughput Encoding of Multi-Kernel Polar Codes}

\author{Hossein Rezaei,~\IEEEmembership{Graduate Student Member,~IEEE},
        Elham Abbasi,~\IEEEmembership{Student Member,~IEEE},
        Nandana Rajatheva,~\IEEEmembership{Senior Member,~IEEE}, and
        Matti Latva-aho,~\IEEEmembership{Senior Member,~IEEE}

\thanks{(Corresponding author: Hossein Rezaei.)
H. Rezaei, N. Rajatheva, and M. Latva-aho are with the Centre for Wireless Communications (CWC),
University of Oulu, 90014 Oulu, Finland (e-mail: hossein.rezaei@oulu.fi;
nandana.rajatheva@oulu.fi; matti.latva-aho@oulu.fi).}}


\markboth{}%
{Shell \MakeLowercase{\textit{et al.}}: A Sample Article Using IEEEtran.cls for IEEE Journals}

\IEEEpubid{0000--0000/00\$00.00~\copyright~2023 IEEE}
\maketitle
\begin{abstract}
Over the past decade, polar codes have received significant traction and have been selected as the coding method for the control channel in fifth-generation (5G) wireless communication systems. However, conventional polar codes are reliant solely on binary ($2\times 2$) kernels, which restricts their block length to being only powers of $2$. In response, multi-kernel (MK) polar codes have been proposed as a viable solution to attain greater code length flexibility.
This paper proposes an unrolled architecture for encoding both systematic and non-systematic MK polar codes, capable of high-throughput encoding of codes constructed with binary, ternary ($3\times 3$), or binary-ternary mixed kernels. The proposed scheme exhibits an unprecedented level of flexibility by supporting $83$ different codes and offering various architectures that provide trade-offs between throughput and resource consumption. The FPGA implementation results demonstrate that a partially-pipelined polar encoder of size $N=4096$ operating at a frequency of $270$ MHz gives a throughput of $1080$ Gbps. Additionally, a new compiler implemented in Python is given to automatically generate HDL modules for the desired encoders. By inserting the desired parameters, a designer can simply obtain all the necessary VHDL files for FPGA implementation.

\end{abstract}
\section{Introduction}
\label{sec_intro}
\IEEEPARstart{A}{r{\i}kan} \cite{Arikan} introduced polar codes as a class of error-correcting codes capable of achieving the symmetric channel capacity of a binary-input discrete memoryless channel (BIDMC) as the code length grows towards infinity. This is achieved through the use of a recursive construction technique, wherein a polar code with size $N=2^n$ can be generated by taking the $n$th Kronecker power of a binary matrix referred to as Ar{\i}kan's kernel. By applying this construction, the physical channel undergoes a transformation into $N$ virtual synthetic channels, with their reliability tending towards either zero or infinity with the growth of the code length.

Ar{\i}kan in \cite{Arikan} also demonstrated that polar codes could achieve the capacity of symmetric channels when utilizing the successive cancellation (SC) decoding algorithm. Since then, researchers have been heavily invested in improving polar codes of restricted size, focusing on parameters such as decoding latency under SC, complexity, power, and error-correction performance.
The introduction of successive cancellation list (SCL) decoding \cite{tal2015list}, coupled with cyclic redundancy check (CRC) \cite{niu2012crc}, has significantly enhanced the error-correction performance of polar codes, allowing them to compete favorably with other channel coding techniques, such as low-density parity-check (LDPC) codes. 
These efforts have paved the way for the adoption of polar codes in the $3$rd generation partnership project (3GPP) fifth-generation new radio ($5$G-NR) wireless communication standard \cite{3GPP}.

Extensive research has been conducted on hardware implementations of polar encoders to cater to diverse application demands. In particular, both feed-forward and feed-back pipelined polar encoders have been proposed in \cite{7169326}. A more comprehensive pipelined encoder architecture that can be used for any code length or parallelism level is presented in \cite{7482677}. This architecture leverages the recursive nature of polar codes, enabling it to be utilized for any code length or parallelism level, thus offering greater flexibility and versatility in practical applications. A folded pipelined encoder is introduced in \cite{6951410}. Building on \cite{6951410}, a pruned folded encoder was developed in \cite{9063643} to achieve higher throughput while reducing complexity. However, the fixed code lengths of $256$ and $1024$ limited its applicability. 
Rather than changing the parallelism level of the encoder, \cite{7482677, 6951410, SHIH2018292} employed different radices for encoding reconfiguration. 
Additionally, first-in first-out (FIFO) modules and pipeline registers were utilized to reduce the critical path of the proposed polar encoder, which enabled high clock frequency and throughput at the expense of higher power consumption \cite{7991021}. A polar compiler is proposed in \cite{Zhong2020} that automatically generates the hardware description of pipelined or stage-folded polar encoders.

\IEEEpubidadjcol
Despite the significant progress made in the development of polar encoders, the current research has primarily concentrated on codes constructed by $2\times2$ polarization matrix  \cite{Arikan}, which limits the block length of polar codes to powers of $2$. As a result, polar codes constructed using Ar{\i}kan's kernel are not capable of addressing all code lengths and rates that are required in modern communication systems beyond 5G networks. The binary Kronecker product has been employed to achieve polarization phenomenon and can be extended to other kernels. The ternary kernel ($3\times3$) is of particular interest due to its low complexity and optimal polarization properties. Multi-kernel (MK) polar codes, which utilize kernels of different dimensions, have leveraged this technique to provide flexible code lengths, making them suitable for modern communication systems. Several MK decoders have been proposed recently \cite{Coppolino, Rezaei2022, rezaei2022combinational, Rezaei2022MK, rezaei2023high}. Nevertheless, to date, research in this field has not addressed the need for a MK encoder, leaving an important gap in the state-of-the-art.

This paper aims to accomplish two key objectives. The primary goal is to present the first-ever systematic and non-systematic MK encoders. The secondary objective is to introduce the cutting-edge SC-based high-throughput architecture for polar encoders. In order to achieve these objectives, we have employed unrolling and pipelining techniques to optimize the performance of systematic and non-systematic MK encoders. This has resulted in hardware frameworks capable of achieving throughput rates in the range of hundreds of Gbps, marking a significant advancement from existing state-of-the-art counterparts. In fact, our achievement represents a remarkable improvement of one to two orders of magnitude in throughput. Additionally, we present a variety of architectures that offer flexible trade-offs between throughput and area, providing a range of options for customization.

In addition, a Python-based hardware compiler is proposed that effortlessly exports all the necessary VHDL files required for the FPGA implementation of the proposed encoders. The idea stems from the fact that altering the block length or kernel ordering dictates modifying all the VHDL modules. However, by entering the necessary parameters to the proposed compiler, all VHDL files will be modified and exported accordingly. Furthermore, in case the user does not enter the kernel ordering, the proposed compiler is capable of calculating the kernel ordering that yields the highest error-correction performance. 

The remainder of this paper are structured as follows. In Section \ref{sec_back}, we provide a comprehensive background on Ar{\i}kan's and MK polar codes, and explicate the polarization performance of various ternary matrices. Section \ref{sec_prop} details the proposed architecture of both systematic and non-systematic MK encoders, while also providing a complexity analysis. This section also presents a comprehensive exposition of the proposed compiler. In Section \ref{sec_result}, we present the implementation results and a comparison with the cutting-edge research in the field. Finally, Section \ref{sec_conc} concludes this work.

\section{Preliminaries}
\label{sec_back}
In this section, alongside offering a comprehensive overview of Ar{\i}kan's and MK polar codes, we present code construction techniques and the encoding process for Ar{\i}kan's, pure-ternary, and MK polar codes. Moreover, we provide a detailed discussion concerning various ternary polarization matrices and their polarization performance.

\IEEEpubidadjcol
\subsection{Ar{\i}kan's Polar Codes}
The symbol $\mathcal{PC}(N,K)$ represents a polar code with a size of $N=2^n$ and $K$ bits of information. The code rate can be computed as $\mathcal{R} = {K/N}$. Ar{\i}kan proved the phenomenon of channel polarization for binary polar codes in \cite{Arikan}. This phenomenon provides the ability to convert a physical communication channel, denoted as $W$, into a set of $N$ distinct virtual channels $\boldsymbol{W}_{i}^{N}$, where $1 \leq i \leq N$. Each virtual channel possesses a unique level of reliability. As $N$ approaches infinity, the reliability of each virtual channel tends towards either perfect reliability ($1$) or perfect unreliability ($0$). To determine which virtual channels are reliable, we can use either the Bhattacharya parameters \cite{Arikan}, or Gaussian approximation \cite{mori2009performance}. The collection of $K$ most reliable bit positions is referred to as the information set, designated by the symbol $\boldsymbol{\mathcal{I}}$, whereas the remaining $N-K$ bit positions constitute the frozen set, denoted by the symbol $\boldsymbol{\mathcal{F}}$. A value of zero is assigned to the frozen bits.

Polar codes can be constructed using a linear transformation expressed as $\boldsymbol{x} = \boldsymbol{u\cdot G}$, where $\boldsymbol{x}$ represents the encoded stream, $\boldsymbol{u}$ is an $N$-bit input vector to the encoder, and $\boldsymbol{G}$ is a squared generator matrix. The input vector $\boldsymbol{u}$ is comprised of reliable and unreliable positions, where the message and frozen data are inserted, respectively. The generator matrix $\boldsymbol{G} = \boldsymbol{T}_2^{\otimes n}$ is formed by taking the n-th Kronecker product (denoted by $\otimes$) of the Ar{\i}kan's kernel, denoted as 
\begin{equation}
\begin{aligned}
\boldsymbol{T}_2=\begin{bmatrix} 1&0\\ 1&1 \end{bmatrix}. 
\label{eq:binmat}
\end{aligned}
\end{equation}
It should be noted that $\boldsymbol{G}$ is constructed recursively, which enables the concatenation of two codes of size $N/2$ to form a polar code of size $N$. The encoded stream within a binary node at level $\lambda$ ($\beta^{b\lambda}$), wherein $N_{\lambda}=2^\lambda$, can be determined by
\begin{equation}
\begin{aligned}
\relax[\beta^{b\lambda}_i, \beta^{b\lambda}_{i + \frac{N_{\lambda}}{2}}] ={} & [\beta^{bl}_i\oplus\beta^{br}_i,\beta^{br}_i],
\label{eq:betak2}
\end{aligned}
\end{equation}
where $i \in [0,\frac{N_{\lambda}}{2}-1]$, and $\oplus$ represents XOR operation. The variables $\beta_i^{bl}$ and $\beta_i^{br}$ denote the data located in the left and right halves of the lower level ($\lambda-1$).
\subsection{MK Polar Codes}
By limiting the generator matrix to the $\boldsymbol{T_2}$ kernel, the block lengths of larger codes are restricted to powers of 2. However, to meet the requirements of potential use cases such as the 5G-NR wireless communication standard \cite{3GPP} and LDPC WiMAX \cite{Shin2012}, code lengths constructed by kernels other than $\boldsymbol{T_2}$ are necessary. By incorporating one or a few non-Ar{\i}kan kernels, we can obtain most of the desired code lengths desired by such standards.
To construct a generator matrix for a code with size $N = l_0\times l_1\times\ldots\times l_s$, we can use a series of Kronecker products between different kernels as
\begin{equation}
    \boldsymbol{G} \triangleq \boldsymbol{T}_{l_0}\otimes \boldsymbol{T}_{l_1}\otimes\ldots\otimes  \boldsymbol{T}_{l_s}.
    \label{eq:genmat}
\end{equation}
The variables denoted by $l_i$ ($0 \leq i \leq s$) are prime numbers that may be identical, while $\boldsymbol{T}_{l_i}$ represents squared matrices. Each unique prime number can serve as a dimension for a kernel. However, the simplest and most appealing non-Ar{\i}kan kernel is the ternary kernel, which can be constructed using three different polarizing matrices \cite{Zhang2012},  as 
\begin{equation}
\begin{aligned}
\boldsymbol{T}_3^1=\begin{bmatrix} 1&0&0 \\ 1&1&0\\ 0&0&1 \end{bmatrix},
~\boldsymbol{T}_3^2=\begin{bmatrix} 1&0&0 \\ 1&1&0\\ 1&0&1 \end{bmatrix},
~\boldsymbol{T}_3^3=\begin{bmatrix} 1&0&0 \\ 0&1&0\\ 1&1&1 \end{bmatrix}.
\label{eq:termat}
\end{aligned}
\end{equation}
The encoded sequence in a ternary node at level $\lambda$ ($\beta^{t\lambda}$), wherein $N_{\lambda}=3^\lambda$, can be determined by
\begin{equation}
\begin{aligned}
\relax[\beta^{t\lambda}_i, \beta^{t\lambda}_{i + \frac{N_{\lambda}}{3}},\beta^{t\lambda}_{i + \frac{2N_{\lambda}}{3}}] ={} & [\beta^{tl}_{i}\oplus\beta^{tc}_{i}, \beta^{tl}_{i}\oplus\beta^{tr}_{i}, \beta^{tl}_{i}\oplus\beta^{tc}_{i}\oplus\beta^{tr}_{i}],
\label{eq:betak3}
\end{aligned}
\end{equation}
where $i \in [0,\frac{N_{\lambda}}{3}-1]$, and the variables $\beta_i^{tl}$, $\beta_i^{tc}$, and $\beta_i^{tr}$ refer to the data residing in the left, middle, and right partitions of the lower level ($\lambda-1$), respectively.

\begin{figure*}
    \centering    \includegraphics[width=2\columnwidth]{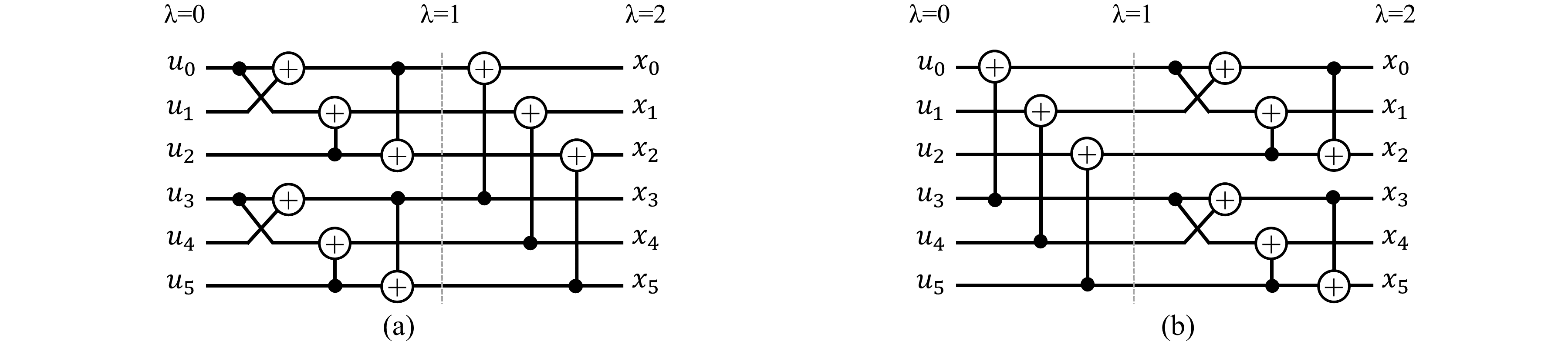}
    \caption{The Tanner graph of a MK polar code of size $N=6$ using a) $\boldsymbol{G}_1 = \boldsymbol{T}_2 \otimes \boldsymbol{T}_3$ and b) $\boldsymbol{G}_2 = \boldsymbol{T}_3 \otimes \boldsymbol{T}_2$.}
    \label{fig:EncDec}
\end{figure*}

In this paper, the codes that are composed of either binary or ternary kernels or a combination of the two will be investigated. This implies that the codes we examine may fall under the category of pure-binary (Ar{\i}kan's), pure-ternary, or binary-ternary mixed polar codes.
The MK codes of this paper have a block length expressed as $N=2^n\cdot 3^m$, where $n$ and $m$ are natural numbers and $0 \leq i \leq n+m-1$. The generator matrix is given by $\boldsymbol{G} = \otimes_{i=0}^{n+m-1} \boldsymbol{T}_{l_i}$.
For instance, let us consider the basic MK polar code of size $N=6$. There exist two distinct kernel sequences, namely $\boldsymbol{G}_1=\boldsymbol{T}_2\otimes \boldsymbol{T}_3$ and $\boldsymbol{G}_2=\boldsymbol{T}_3\otimes \boldsymbol{T}_2$, leading to two different generator matrices. This occurs because the Kronecker product is non-commutative. Therefore, unique polar codes with different performance characteristics are formed depending on the kernel orderings. Fig. \ref{fig:EncDec} presents the Tanner graphs of the MK code of size $N=6$ corresponding to two generator matrices of $\boldsymbol{G}_1$ and $\boldsymbol{G}_2$.
\subsection{Comparison of Different Ternary Constituent Matrices and Measure of Polarization Performance}
It is shown in \cite{Zhang2012} that all polarization matrices can be classified into three distinct categories, each of which serves to identify the positions of information and frozen bits. Moreover, there exist specific matrices that exhibit identical unordered Bhattacharyya parameters. Thus it is rational to group them together within the same classification. An overview of these three classifications corresponding to the matrices presented in (\ref{eq:termat}) is provided in Table \ref{tab:classification}.

\begin{figure}
    \centering    \includegraphics[width=1\columnwidth]{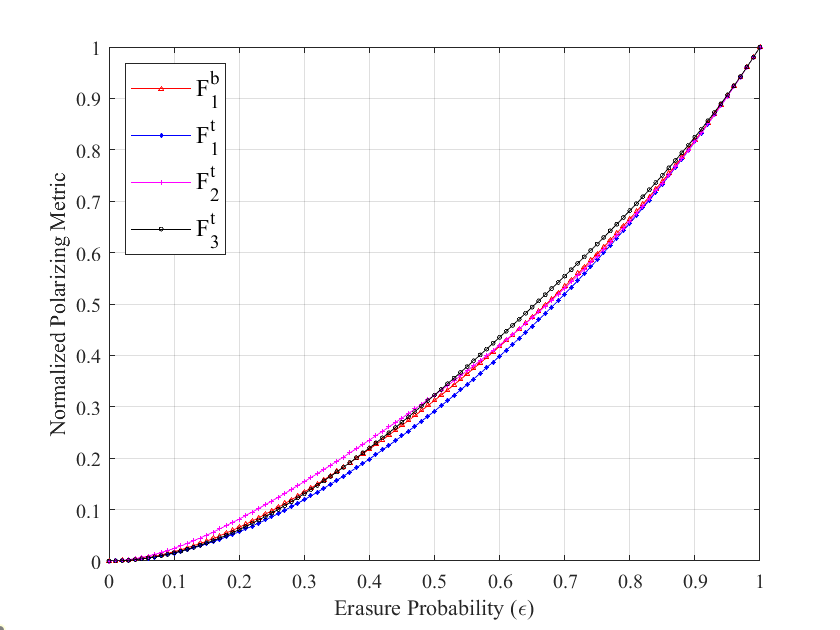}
    \caption{Measure of normalized polarization performance for different erasure probabilities.}
    \label{fig:Bhattacharyya1}
\end{figure}

\begin{figure}
    \centering    
    \includegraphics[width=1\columnwidth]{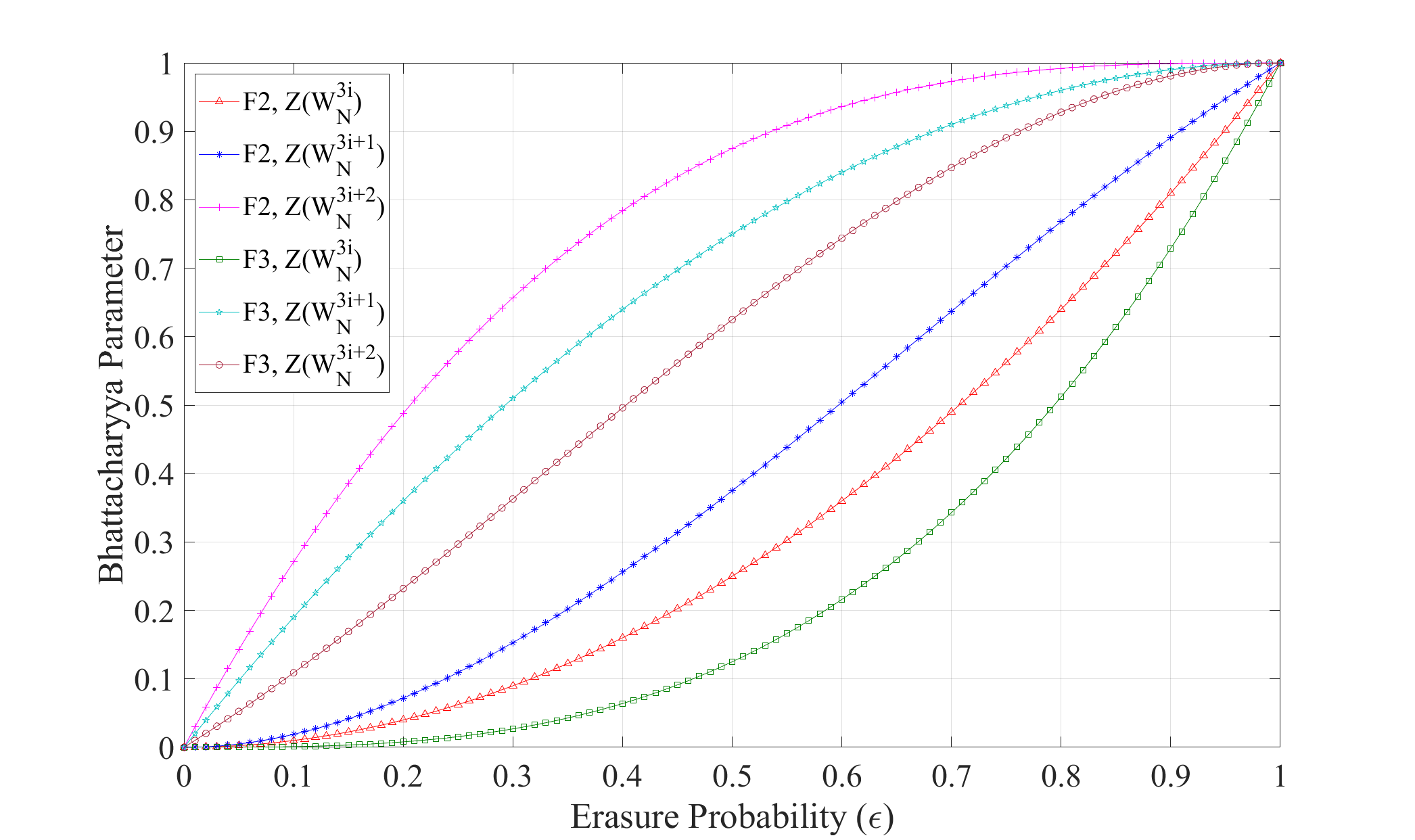}
    \caption{Bhattacharyya parameter evolution for a single ternary stage.}
    \label{fig:Bhattacharyya2}
\end{figure}

In \cite{8746303}, a parameter denoted as $M(\boldsymbol{F}_l)$ is defined as a measure of polarization performance of a desired kernel $l$ as
\begin{equation}
    M(\boldsymbol{F}_l)=\Sigma_{i=0}^{l-1} \boldsymbol{Z}^2(W_l ^i).
    \label{eq:Polmetric}
\end{equation}
The higher value of $M(\boldsymbol{F}_l)$ corresponds to an improved degree of polarization.
The measure of classification for binary and three distinct ternary matrices of Table \ref{tab:classification} can be computed as 
\begin{table}
\centering
\caption{Classification of unordered Bhattacharyya parameters for ternary matrices}
\begin{tabular}{c|ccc}
\hline
Formula&\multicolumn{3}{c}{Bhattacharyya parameters}\\\hline
F$^t_1$&$\epsilon$&$\epsilon ^2$&$-\epsilon ^2+2\epsilon$\\ \hline
F$^t_2$&$\epsilon ^2$&$-\epsilon ^3+2\epsilon^2$&$\epsilon^3-3\epsilon ^2+3\epsilon$\\\hline
F$^t_3$&$\epsilon ^3$&$-\epsilon ^2+2\epsilon$&$-\epsilon^3+\epsilon ^2+\epsilon$\\
\hline
\end{tabular}
\label{tab:classification}
\end{table}
\begin{equation}
    M(\boldsymbol{F}^b_1)=\frac{1}{2}(2\epsilon^4-4\epsilon^3+4\epsilon^2),
    \label{eq:Polmetricg2}
\end{equation}
\begin{equation}
    M(\boldsymbol{F}_1^{t})=\frac{1}{3}(2\epsilon^4-4\epsilon^3+5\epsilon^2),
    \label{eq:Polmetricg31}
\end{equation}
\begin{equation}
    M(\boldsymbol{F}_2^{t})=\frac{1}{3}(2\epsilon^6-10\epsilon^5+20\epsilon^4-18\epsilon^3+9\epsilon^2),
    \label{eq:Polmetricg32}
\end{equation}
\begin{equation}
    M(\boldsymbol{F}_3^{t})=\frac{1}{3}(2\epsilon^6-2\epsilon^5-2\epsilon^3+5\epsilon^2).
    \label{eq:Polmetricg33}
\end{equation}
Equations (\ref{eq:Polmetricg2})-(\ref{eq:Polmetricg33}) are compared in Fig. \ref{fig:Bhattacharyya1}. It is evident that $\boldsymbol{F}_1^{t}$ displays the poorest performance, while $\boldsymbol{F}_2^{t}$ and $\boldsymbol{F}_3^{t}$ exhibit superior results compared to $\boldsymbol{F}_1^{b}$. Additionally, it is notable that for values of $\epsilon \in (0, 0.5)$, $\boldsymbol{F}_2^{t}$ demonstrates the most outstanding performance, whereas for values of $\epsilon \in (0.5, 1)$, $\boldsymbol{F}_3^{t}$ yields the highest performance.
Fig. \ref{fig:Bhattacharyya2} illustrates the evolution of all ternary indices related to $\boldsymbol{F}_2^{t}$ and $\boldsymbol{F}_3^{t}$. We can obviously observe that 
\begin{equation}
\begin{aligned}
Z(W_N^{3i+2})\geq Z(W_N^{3i+1})\geq Z(W_N^{3i}).
\label{eq:ZW}
\end{aligned}
\end{equation}
Hence, it can be inferred that index $3i$ stochastically experiences capacity degradation while indices $3i+1$ and $3i+2$ undergo enhancement in capacity. In this paper, we utilize the $\boldsymbol{F}_3^{t}$ formula due to higher polarization performance at indices $3i+1$ and $3i+2$ compared to that of index $3i$. Moreover, the aforementioned indices display significantly less disparity in contrast to those of $\boldsymbol{F}_2^{t}$.

\begin{figure*}
    \centering    \includegraphics[width=2\columnwidth]{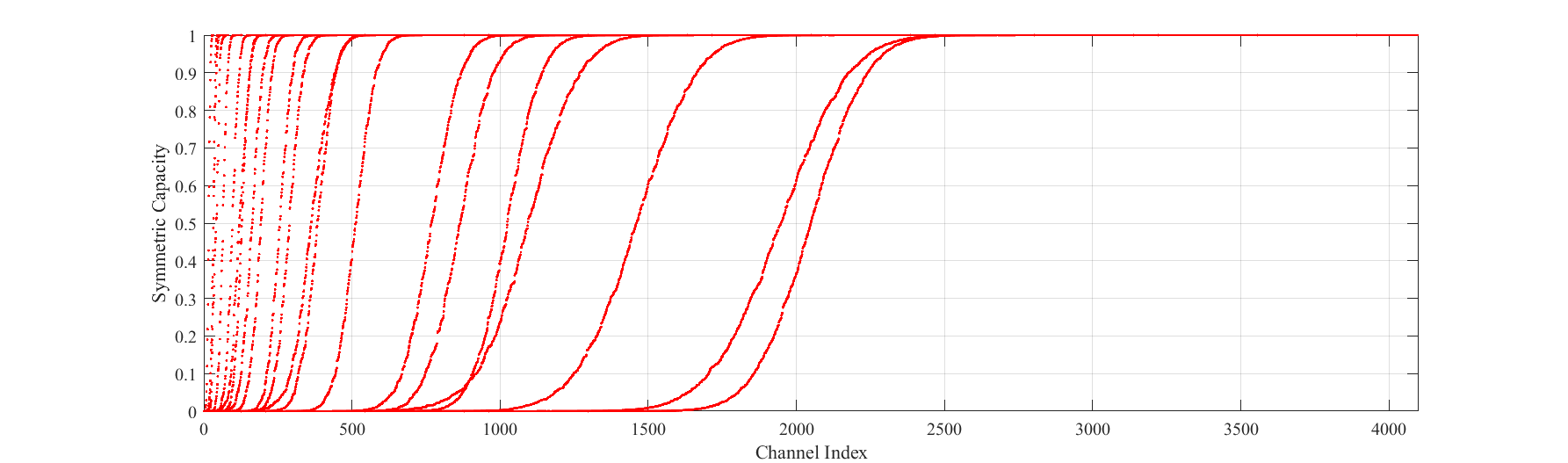}
    \caption{Symmetric capacity versus channel index of different pure-binary, pure-ternary, and MK codes with $\epsilon=0.5$.}
    \label{fig:basicblocks}
\end{figure*}

The symmetric channel capacity of multiple pure-binary, pure-ternary, and MK polar codes over an erasure channel with an erasure probability of $\epsilon = 0.5$ is depicted in Fig. \ref{fig:basicblocks}. Obviously, the behavior of pure-ternary and MK codes is similar to that of Ar{\i}kan's codes, albeit with a potentially lower degree of polarization performance. It can be seen that as the indices become smaller, the symmetric capacity approaches zero, whereas for larger indices, it approaches one. However, an unpredictable pattern can be noticed for indices within the intermediate range.
\section{The Proposed Architectures of MK Polar Encoders}
\label{sec_prop}
As previously noted, the performance of a MK encoder is highly dependent on the kernel ordering, and constantly changing it can lead to sub-optimal encoding performance. Furthermore, the Kronecker product represents a non-commutative operation, which complicates the design of a generalized semi-parallel MK encoder capable of handling a dynamically changing kernel order. This task is exceptionally challenging due to the complexity inherent in non-commutative operations. 
Since the SC algorithm does not include any loops, it is possible to develop an unrolled architecture that does not require memory elements between the input and output stages. Given these considerations and our primary objective of achieving high throughput, we propose utilizing unrolled architecture for MK encoders. Moreover, the encoder's throughput can be further enhanced by implementing pipelining at any depth.

This section provides a description of the proposed non-systematic and systematic combinational architecture of Ar{\i}kan's, pure-ternary and binary-ternary mixed polar codes, followed by an overview of the memory architecture in the proposed SC-based MK encoders.

\subsection{Non-Systematic Unrolled Combinational MK Architecture}
\begin{figure*}
    \centering
    \includegraphics[width=1.5\columnwidth]{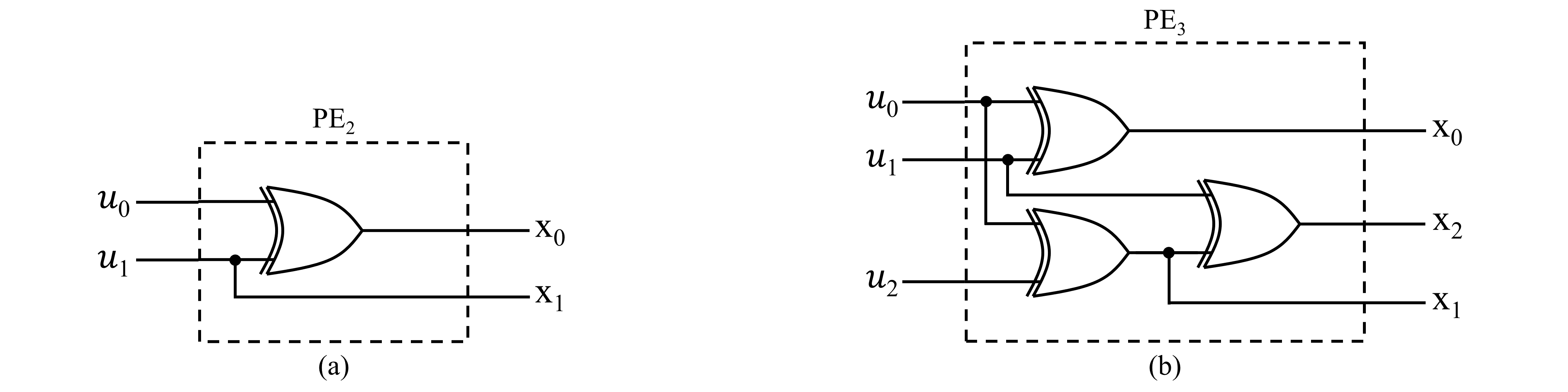}
    \caption{The basic PEs of (a) Ar{\i}kan's and (b) pure-ternary encoders corresponding to polar codes of sizes $N=2$ and $N=3$, respectively.}
    \label{fig:BinDecLog}
\end{figure*}

\begin{figure}
    \centering
    \includegraphics[width=1\columnwidth]{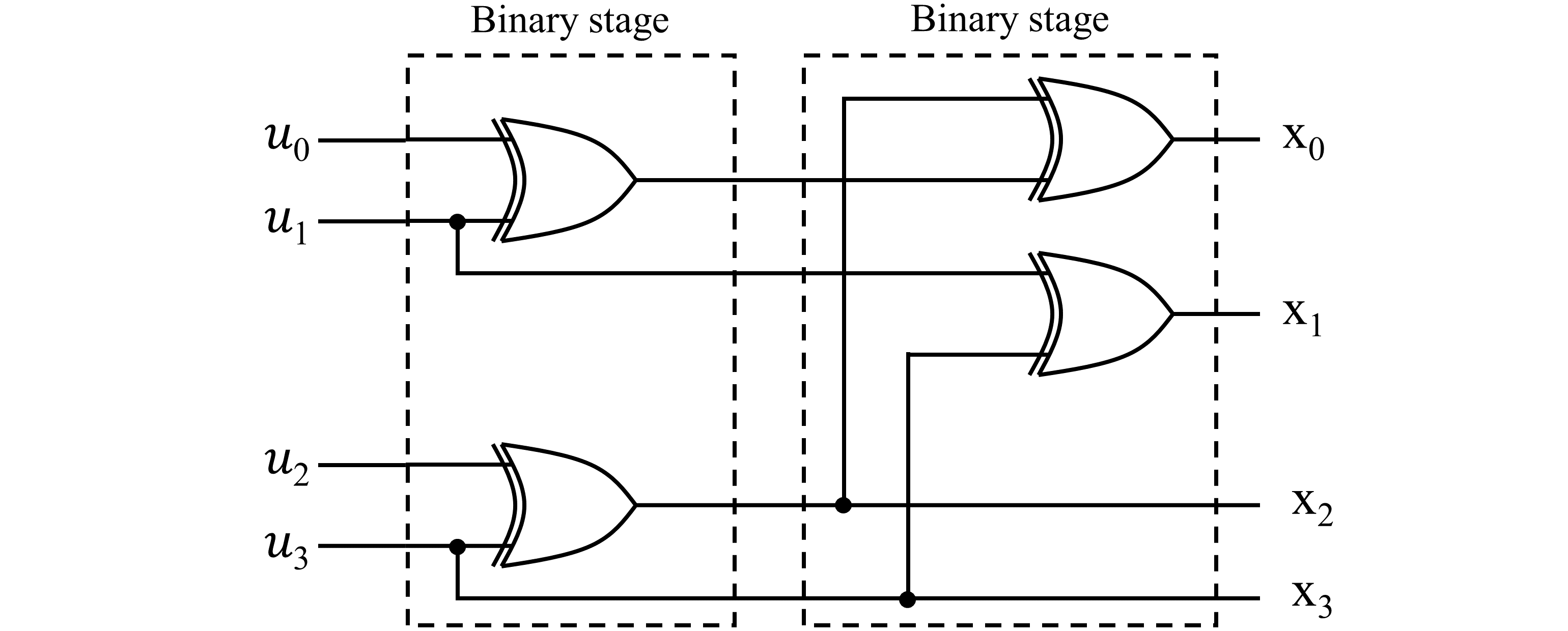}
    \caption{The pure-combinational non-systematic Ar{\i}kan's encoder of size $N=4$.}
    \label{fig:RecConstT2}
\end{figure}

\begin{figure}
    \centering
    \includegraphics[width=1\columnwidth]{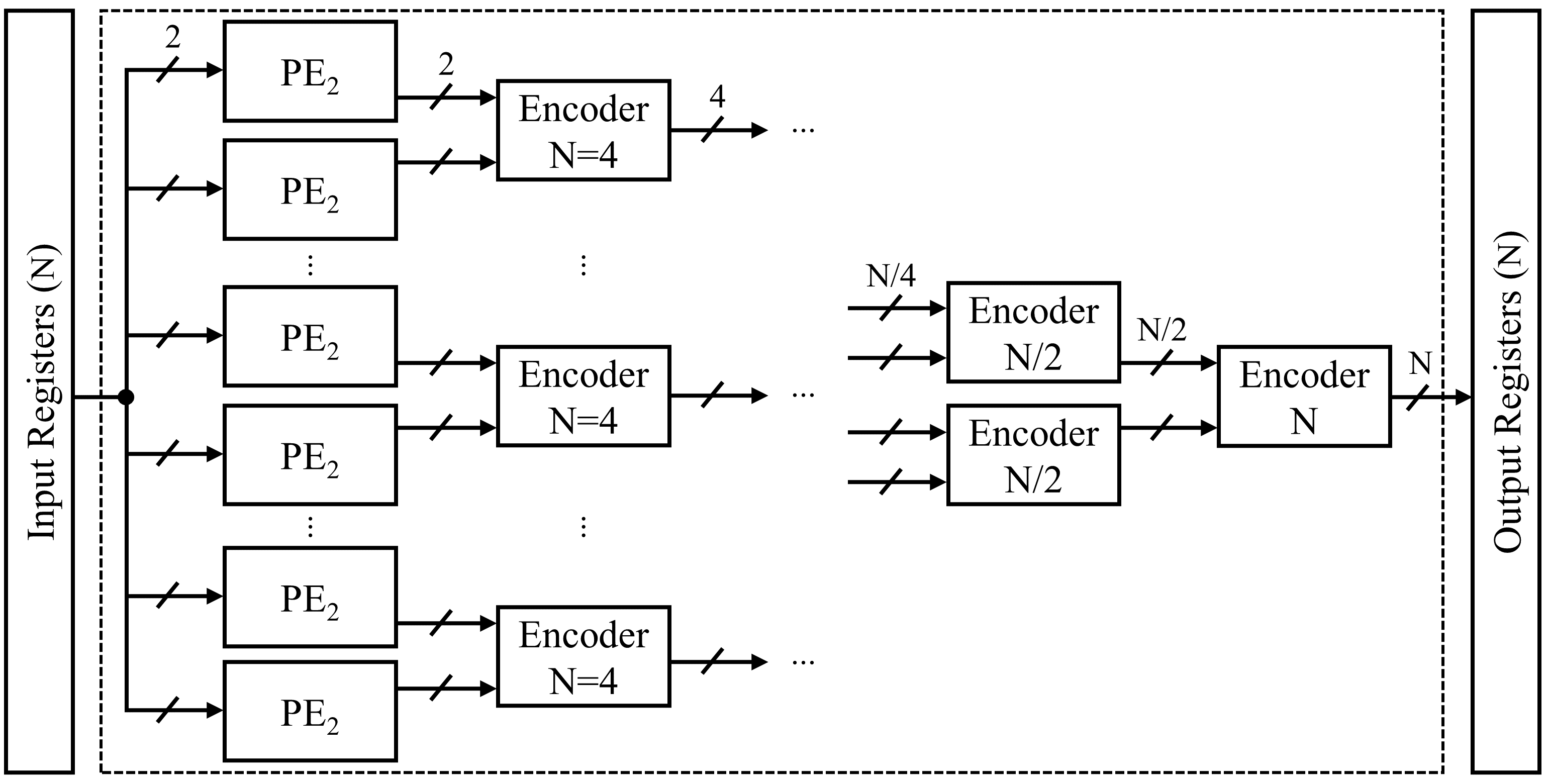}
    \caption{The pure-combinational non-systematic Ar{\i}kan's encoder of size $N$.}
    \label{fig:K2Datapath}
\end{figure}

\begin{figure}
    \centering
    \includegraphics[width=1\columnwidth]{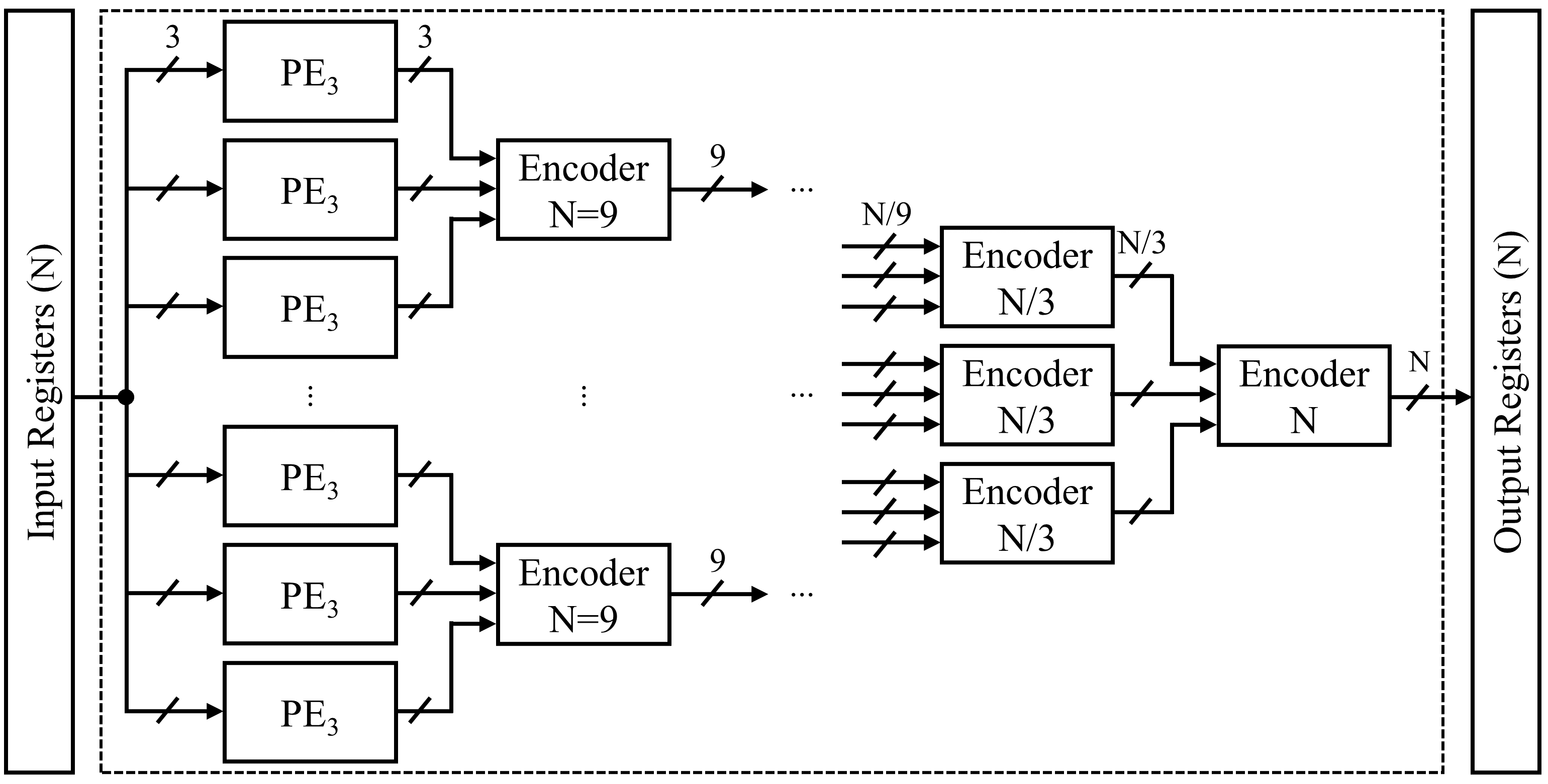}
    \caption{The pure-combinational non-systematic ternary encoder of size $N$.}
    \label{fig:K3Datapath}
\end{figure}

\begin{figure}
    \centering
    \includegraphics[width=1\columnwidth]{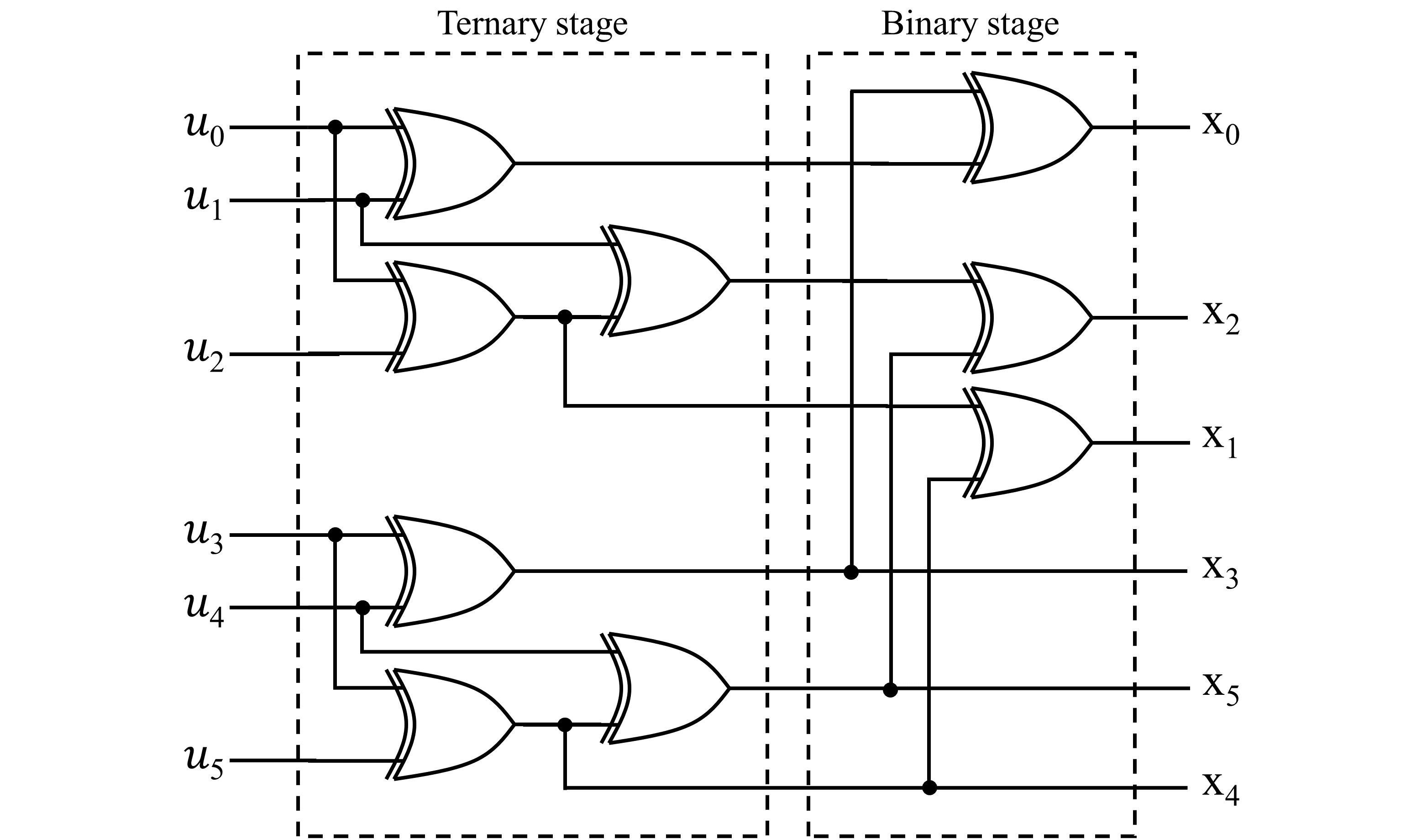}
    \caption{The pure-combinational non-systematic MK encoder of size $N=6$ using the kernel ordering as $\boldsymbol{G} = \boldsymbol{T}_2 \otimes \boldsymbol{T}_3$.}
    \label{fig:RecConstMK}
\end{figure}

Fig. \ref{fig:BinDecLog} depicts the processing elements (PEs) utilized by Ar{\i}kan's as well as that of the pure-ternary polar encoders. These PEs correspond to polar codes of sizes $N=2$ and $N=3$, respectively. The propagation delay of Ar{\i}kan's and ternary PEs can be denoted as $t_{pd}^b$ and $t_{pd}^t$, respectively. By considering the propagation delay of an XOR gate as $t_{pd}^{XOR}$, the $t_{pd}^b$ and $t_{pd}^t$ can be estimated as
\begin{equation}
\begin{aligned}
t_{pd}^b = t_{pd}^{XOR},\
t_{pd}^t = 2 \cdot t_{pd}^{XOR}.
\label{eq:tpdt}
\end{aligned}
\end{equation}
Hence, it can be observed that the propagation delay of the ternary PE is twice as long as that of Ar{\i}kan's PE. In other words, mathematically expressed as 
\begin{equation}
\begin{aligned}
t_{pd}^t = 2 \cdot t_{pd}^b. 
\label{eq:tpdTvsB}
\end{aligned}
\end{equation}
As shown in Fig. \ref{fig:RecConstT2}, the utilization of two stages of Ar{\i}kan's PE (indicated by dashed boxes) can construct a non-systematic encoder for a polar code of size $N=4$. By applying this approach recursively, we can extend the encoder to a larger size of $N=2^n$ by utilizing $n$ consecutive stages of Ar{\i}kan's PE (PE$_2$ as illustrated in Fig. \ref{fig:BinDecLog} (a)). 
The recursive architecture of the pure-combinational non-systematic Ar{\i}kan's encoder of size $N$ is illustrated in Fig. \ref{fig:K2Datapath}.
Following the same methodology as Ar{\i}kan's encoder, it is possible to construct a pure-ternary non-systematic encoder of size $N=3^m$ in a recursive manner, exploiting $m$ consecutive stages of the ternary PE (PE$_3$ as illustrated in Fig. \ref{fig:BinDecLog} (b)). The overall recursive architecture of such encoders is depicted in Fig. \ref{fig:K3Datapath}. 
By using both Ar{\i}kan's and ternary PEs, we can construct an architecture for a MK polar code of size $N=2^n\cdot 3^m$. The pure-combinational architecture of a non-systematic polar encoder of size $N=6$, with kernel ordering of $\boldsymbol{G} = \boldsymbol{T}_2 \otimes \boldsymbol{T}_3$, is portrayed in Fig. \ref{fig:RecConstMK}.

The proposed architectures contains two sets of registers, each with a size of $N$ bits, which serve to store the input data ($N$ bits) and the encoded codeword ($N$ bits). A close inspection of Fig. \ref{fig:K2Datapath} and Fig. \ref{fig:K3Datapath} reveal that the non-systematic combinational encoders, regardless of their kernel orderings, lack any synchronous logic components, such as registers or RAM arrays, between the input and output registers. This characteristic of combinational encoders results in higher power and processing time efficiency. Furthermore, the absence of RAM routers reduces hardware complexity and eliminates long read/write latencies. The combinational encoder's latency is limited to one clock cycle as it generates the estimated codeword in a single long clock cycle after accepting the input stream. Therefore, the logic circuitry between the input and output registers determines the overall critical path of the system.

\subsection{Systematic Unrolled Combinational MK Architecture}
The conventional form of polar codes is non-systematic, which means that the information bits are not directly accessible in the encoded codeword. As a result, to retrieve the information bits, a bit-reversal operation is required, which can increase the decoding complexity and introduce additional processing delay. On the other hand, systematic polar codes, as described in \cite{arikan2011systematic}, include the information bits as an integral part of the codeword, thereby eliminating the need for bit-reversal operations. The systematic polar codes offer simpler decoding, improved error-correction performance (as shown in \cite{arikan2011systematic}), and compatibility with legacy systems.

\begin{figure*}
    \centering    \includegraphics[width=1.9\columnwidth]{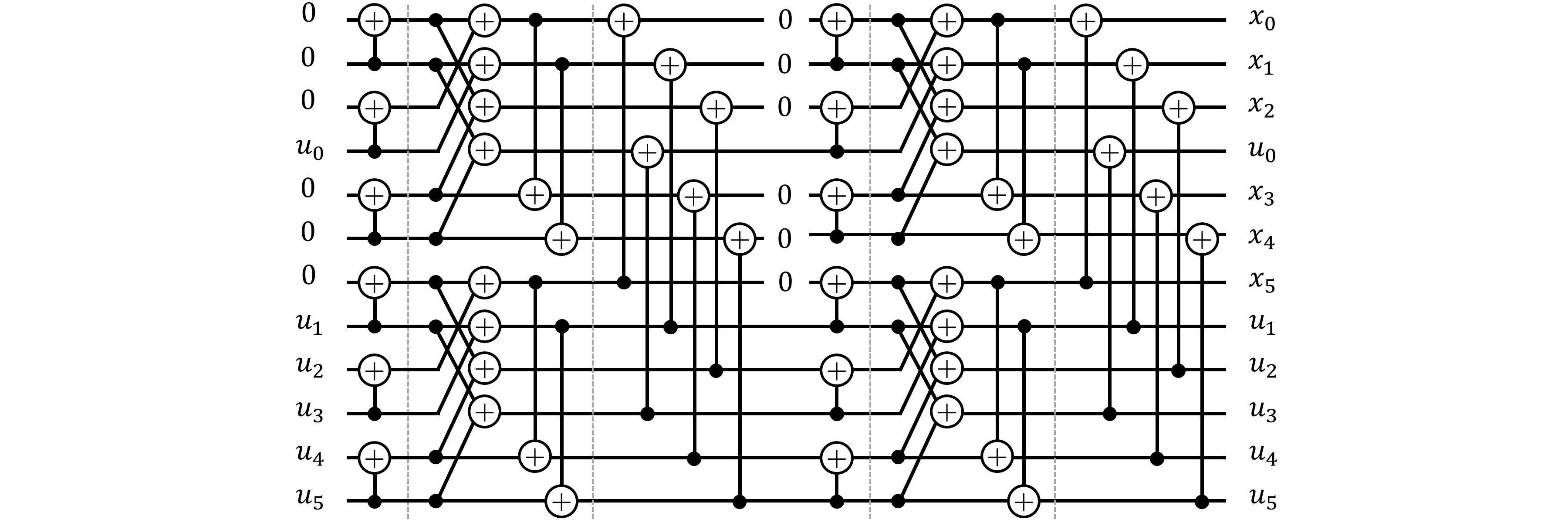}
    \caption{The Tanner graph of a systematic MK polar code of $\mathcal{PC}(12,6)$ using $\boldsymbol{G}_1 = \boldsymbol{T}_2 \otimes \boldsymbol{T}_3 \otimes \boldsymbol{T}_2$.}
    \label{fig:EncDecSys}
\end{figure*}

A systematic encoder of size $N$ can be constructed by connecting two non-systematic encoders of the same size with a zeroing operation applied to the frozen positions between them. For instance, Fig. \ref{fig:EncDecSys} displays a systematic encoder of a code of size $N=12$ that carries $K=6$ bits of information. 
This approach allows any advancements in non-systematic polar code encoding to be directly applied to systematic encoding. Therefore, in the case of a systematic encoder, one can anticipate a greater degree of latency and resource utilization, coupled with a lower throughput when compared to its non-systematic counterpart.

\begin{figure}
    \centering    \includegraphics[width=1.1\columnwidth]{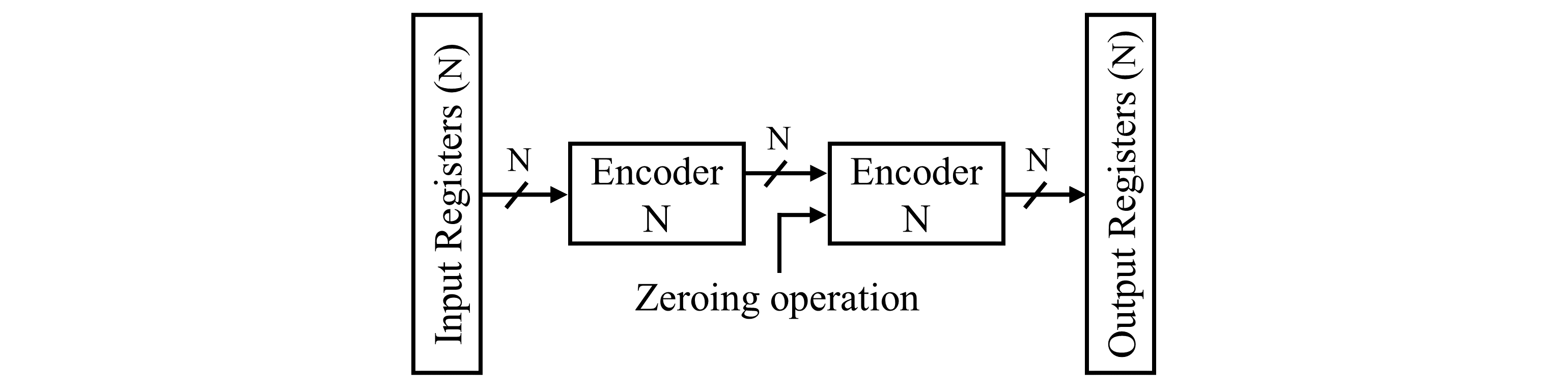}
    \caption{The overall architecture of the proposed unrolled combinational systematic encoder of size $N$.}
    \label{fig:SysArch}
\end{figure}

The overall architecture of the unrolled combinational systematic encoder with a size of $N$ is depicted in Fig. \ref{fig:SysArch}. As can be seen from the figure, the encoder is sandwiched between two sets of registers, each with a size of N.

\subsection{The Proposed Architecture of MK Pipelined Encoders}
The systematic and non-systematic unrolled combinational architectures are associated with a high latency, which can limit their efficiency in real-life application scenarios that demand high encoding throughput. In such cases, the feed-forward pipelined architecture presents a more suitable solution, enabling faster and more efficient encoding processes.
In the case of a pipelined encoder, the amount of memory usage is significantly larger in order to persist data. The input and output memories are identical to those of the combinational encoders. The memory usage increases by a factor of $N$ with the addition of each pipeline stage to the encoder. However, the encoder can operate at a higher clock frequency when implemented with pipelining, and the depth of the pipeline can be adjusted as necessary. Therefore, when designing a pipelined encoder, it is important to consider the trade-off between memory usage and clock rate. 
In the case of a deeply-pipelined encoder, it is possible to load a new data frame and generate an encoded output at every clock cycle. In pipelined encoders, it is typically the case that the number of frames being encoded at any given cycle is equal to the number of pipeline stages plus one ($\mathcal{P}+1$). The result is an extremely high throughput encoder at the cost of high memory requirements. 

When designing pipelined encoders, it is crucial to consider that the combinational logic delay between every pair of pipeline stages must be twice the register's delay or more for optimal performance. As a general rule, it is recommended to incorporate a minimum of two XOR gates, where the propagation delay is calculated as twice the XOR gate's individual propagation delay ($t_{pd}=2 \cdot t_{pd}^{XOR}$), between a pair of pipeline stages. This condition is inherently fulfilled in the ternary stages, as the minimum propagation delay corresponds to the delay of two XOR gates, as stated in equation (\ref{eq:tpdt}). However, in the case of Ar{\i}kan's stages, it is necessary to concatenate a minimum of two stages consecutively to comply with this criterion. Based on the results obtained from our simulation, we have observed that a polar code with a deep pipeline architecture, having $\mathcal{P}=9$ and $N=1024$, delivers an equivalent throughput compared to its counterpart that employs a pipeline of $\mathcal{P}=4$ stages. It is worth noting that the former architecture requires twice the resources of the latter.

\begin{figure}
    \centering    \includegraphics[width=1.1\columnwidth]{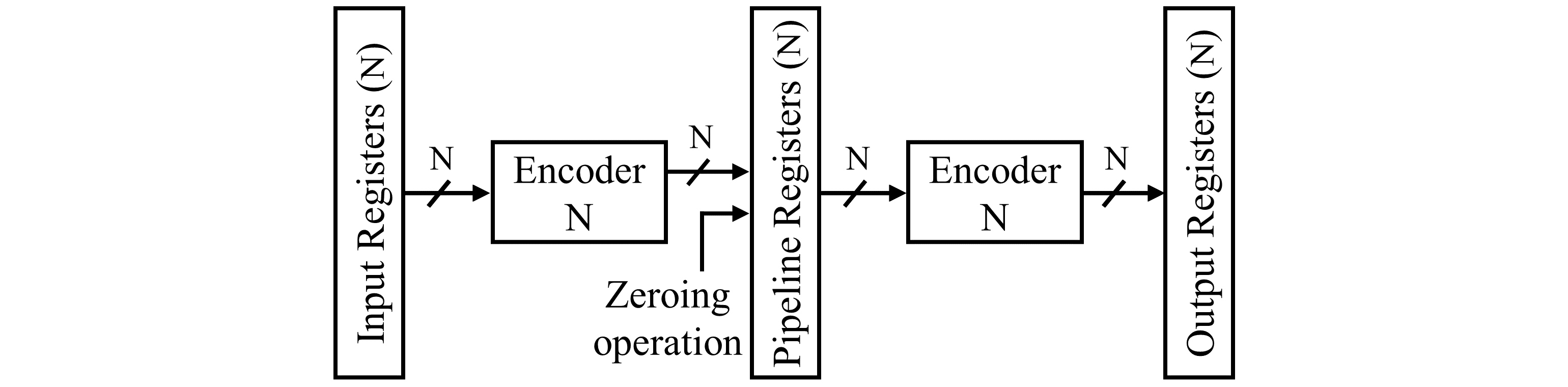}
    \caption{The overall architecture of the proposed unrolled systematic encoder of size $N$ with a pipeline stage at the stage boundaries.}
    \label{fig:SysArchPip}
\end{figure}

In the case of systematic encoders, a specific approach to pipelining involves the incorporation of a pipeline stage at the interface between the two non-systematic encoders. The architecture depicted in Fig. \ref{fig:SysArchPip} exhibits an 
admirable feature, wherein its throughput is approximately equal to that of its non-systematic counterpart.
\subsection{Complexity Analysis of MK Encoders}
The complexity of the proposed polar encoder architecture can be measured by the sum of its registers and XOR gates, similar to \cite{rezaei2023high}. Throughout the paper, all encoders are placed between the input and output registers. With the assumption of using $\mathcal{P}$ pipeline stages, a non-systematic encoder of size $N$ will utilize $r^b_N=(\mathcal{P}+2)\cdot N$ registers.

Now, Let's assume that a non-systematic Ar{\i}kan's encoder of size $N$ consumes $M^b_N$ XOR gates. From the recursive nature of the SC algorithm, it can be inferred that
\begin{equation}
M_N^b = 2M_{\frac{N}{2}}^b+\frac{N}{2}=2(2M_{\frac{N}{4}}^b+\frac{N}{4})+\frac{N}{2}=\ldots. 
\label{bXOR_Gates}
\end{equation}
By initializing equation (\ref{bXOR_Gates}) with $M_2=1$, the total number of XOR gates used in the encoder can be precisely calculated as $\frac{N}{2}\log_2N$. As a result, the total number of basic logic blocks with comparable complexity can be estimated as follows.
\begin{equation}
M_N^b+r_N^b = N\cdot(\frac{1}{2}\log_2N+\mathcal{P}+2).
\label{complex_bin}
\end{equation}
This proves that he complexity of a non-systematic pure-combinational architecture with input and output registers falls within the order of $N\cdot\log_2N$.

In the case of non-systematic pure-ternary encoders, the total number of required XOR gates can be calculated as bellows.
\begin{equation}
M_N^t = 3M_{\frac{N}{3}}^t+\frac{4N}{3}=3(3M_{\frac{N}{9}}^t+{\frac{4N}{9}})+\frac{4N}{3}=\ldots.
\label{tXOR_Gates}
\end{equation}
Knowing that the number of XOR gates used in a ternary encoder of size $N=3$ is $M_3^t = 4$, the exact solution of equation (\ref{tXOR_Gates}) is $\frac{4N}{3}\log_3N$. 
Thus, the total number of the basic logic blocks of a non-systematic pure-ternary encoder can be given by
\begin{equation}
M_N^t+r_N^t = N\cdot(\frac{4}{3}\log_3N+\mathcal{P}+2),
\label{complex_ter}
\end{equation}
which verifies that the complexity of a pure combinational ternary encoder is in the order of $N\cdot\log_3N$. 

When comparing non-systematic Ar{\i}kan's polar codes with similar block lengths to pure-ternary codes, such as $N^b=2048$ and $N^t=2187$, it becomes apparent that Ar{\i}kan's codes have lower complexity. This is because ternary codes have more complex encoding rules than Ar{\i}kan's codes. The complexity of the non-systematic MK encoders ($\mathcal{O}_N^{MK}$) proposed in this paper, with $N_{min}=2$ and $N_{max}=32768$, falls within the lower bound of $N\cdot(\frac{1}{2}\log_2N+\mathcal{P}+2)$ and the upper bound of $N\cdot(\frac{4}{3}\log_3N+\mathcal{P}+2)$, i.e. 
\begin{equation}
N\cdot(\frac{1}{2}\log_2N+\mathcal{P}+2) \leq \mathcal{O}_N^{MK} \leq N\cdot(\frac{4}{3}\log_3N+\mathcal{P}+2).
\label{complex_MK}
\end{equation}
It is worth noting that it is not possible to derive a general equation for the complexity analysis of the MK encoders because it directly depends on the number and order of various kernels in the kernel sequence.

In the case of systematic encoders, the complexity can be determined through the same methodology. The complexity of a systematic combinational Ar{\i}kan's encoder can be computed as
\begin{equation}
M_N^{bs}+r_N^{bs} = N\cdot(\log_2N+2\mathcal{P}+3).
\label{complex_binSys}
\end{equation}
Furthermore, it can be easily proved that the complexity of a systematic combinational pure-ternary encoder can be calculated using the following equation.
\begin{equation}
M_N^{ts}+r_N^{ts} = N\cdot(\frac{8}{3}\log_3N+2\mathcal{P}+3).
\label{complex_terSys}
\end{equation}
Lastly, the complexity of a systematic MK encoder, which has a size of $N$ and satisfies the condition $N_{min}=2$ and $N_{max}=32768$, lies within the lower bound of $N\cdot(\log_2N+2\mathcal{P}+3)$ and the upper bound of $N\cdot(\frac{8}{3}\log_3N+2\mathcal{P}+3)$. In other words, the complexity of a systematic MK encoder ($\mathcal{O}_N^{MKs}$) can be expressed mathematically as follows.
\begin{equation}
N\cdot(\log_2N+2\mathcal{P}+3) \leq \mathcal{O}_N^{MKs} \leq N\cdot(\frac{8}{3}\log_3N+2\mathcal{P}+3).
\label{complex_MKSys}
\end{equation}

\subsection{Automatic Process of Generating HDL Files of Combinational Polar Encoders}
High-level synthesis (HLS) is a process whereby a hardware description of an algorithm or behavior is automatically generated from a high-level programming language, such as Python, SystemC, or C/C++, without requiring significant manual intervention, as defined by Nane et al. in their 2015 survey \cite{nane2015survey}. To generate a register-transfer level (RTL) implementation, HLS tools utilize a hardware description language (HDL), such as Verilog or VHDL. HLS has been increasingly adopted in recent years due to its ability to abstract away low-level design details. Consequently, this paper proposes an automatic hardware generator to expedite the development process of MK encoders, as altering either the architecture or the code characteristics necessitates changing all the HDL sources.

\begin{figure}
    \centering    \includegraphics[width=1\columnwidth]{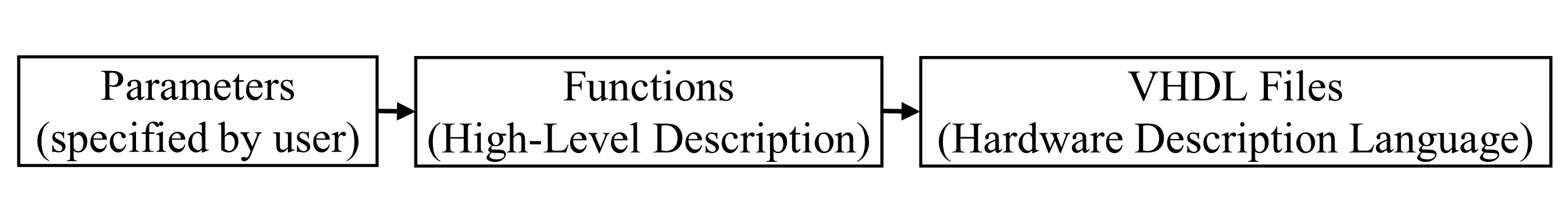}
    \caption{Auto-generation of hardware architectures for target polar encoders.}
    \label{fig:Gen_Process}
\end{figure}

\begin{figure}
    \centering
    \includegraphics[width=1\columnwidth]{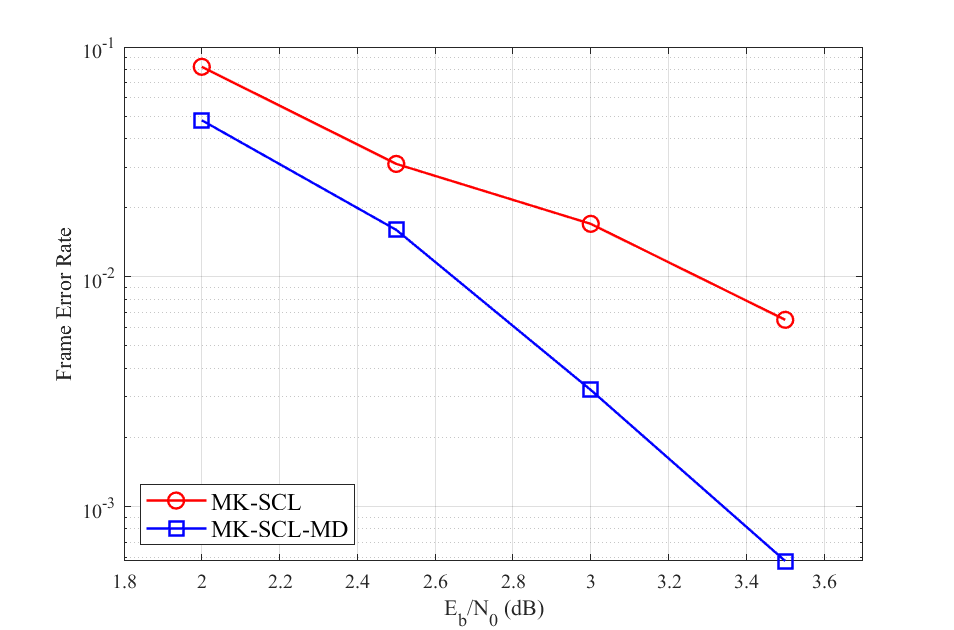}
    \caption{The error-correction capabilities of $\mathcal{PC}(144,72)$, constructed using the methods described in \cite{gabry2017} (referred to as MK-SCL) and \cite{bioglio} (referred to as MK-SCL-MD).}
    \label{fig:MKCompSCL2}
\end{figure}

The proposed compiler \cite{PythonCode}, written in Python, employs multiple functions that correspond to different architectures with user-specified parameters. It employs predefined rules to call relevant functions and facilitate the process of compilation. The functions obtain their inputs from the compiler's top module and generate the desired VHDL files as output. In general, the process of compilation displays certain similarities to the HLS flow as shown in Fig. \ref{fig:Gen_Process}.
To use this tool, the user only needs to provide the block length, the intended configuration of hardware architecture and, if desired, a prefered kernel ordering for the target polar encoder. If a kernel order is not specified, the compiler selects the kernel ordering with the highest error-correction performance using the method proposed in \cite{bioglio}. Fig. \ref{fig:MKCompSCL2} illustrates a comparison of the error-correction performance between the methods presented by \cite{bioglio} and \cite{gabry2017}. The results demonstrate that the method proposed by \cite{bioglio} exhibits superior performance in error-correction.

Algorithm \ref{Alg:SW} provides a comprehensive and detailed depiction of the software's execution sequence, offering a clear and concise understanding of the underlying concept behind the execution process. In Algorithm \ref{Alg:SW}, below notations are used. \\
\textbullet~$sys$: Systematic encoder indicator\\
\textbullet~$T_c$: CPU's elapsed time\\
\textbullet~$N\_temp$: Temporary block length\\
\textbullet~$pipelined$: Pipelined encoder indicator\\
\textbullet~$pip\_depth$: Pipeline depth\\
\textbullet~$pipln\_bndry$: The stage boundaries piplining indicator\\
\textbullet~$\textbf{\emph{ker\_ordr}}$: The kernel ordering\\
\textbullet~$\textbf{\emph{NPS}}$: Specifies the size of codes that require pipeline stages.\\
\textbullet~\emph{PE2\_Func(.)}: Exports the Ar{\i}kan's PE module\\
\textbullet~\emph{PE3\_Func(.)}: Exports the ternary PE module\\
\textbullet~\emph{Pip2Enc\_Func(.)}: Exports the pipelined Ar{\i}kan's sub-module for a code of size $N\_temp$\\
\textbullet~\emph{Comb2Enc\_Func(.)}: Exports the combinational Ar{\i}kan's sub-module for a code of size $N\_temp$\\
\textbullet~\emph{Pip3Enc\_Func(.)}: Exports the pipelined ternary sub-module for a code of size $N\_temp$\\
\textbullet~\emph{Comb3Enc\_Func(.)}: Exports the combinational ternary sub-module for a code of size $N\_temp$\\
\textbullet~\emph{CombEnc\_Reg\_Func(.)}: Exports the top level module in non-systematic encoders containing input and output registers\\
\textbullet~\emph{CombSysEnc\_Reg\_Func(.)}: Exports the top level module in systematic encoders containing input and output registers\\
\textbullet~\emph{PipSysEnc\_Reg\_Func(.)}: Exports the top level module in systematic encoders containing registres in input, output and stage boundaries
\begin{algorithm}
\textbf{Inputs}:  $N$, $sys$, $pipelined$, $pip\_depth$, $\textbf{\emph{ker\_ordr}}$\\
\textbf{Outputs}: VHDL modules, $T_c$\\

\If{$\textbf{ker\_ordr}$ is empty}
{  \textbf{Compute} $\textbf{\emph{ker\_ordr}}$}
\If{$\textbf{pipelined}$ is 1}
{  \textbf{Compute} $\textbf{\emph{NPS}}$ using $pip\_depth$}
\eIf{$\textbf{ker\_ordr}[\textbf{len}(\textbf{ker\_ordr})-1]$ is 2}
{\emph{PE2\_Func()}}{\emph{PE3\_Func()}}
\For{i in \textbf{len}(\textbf{\emph{ker\_ordr}})-2 \textbf{to} $0$ \textbf{by} $-1$}{   
    $N\_temp~*=~\textbf{\emph{ker\_ordr}}[i]$\\
    \eIf{$\textbf{ker\_ordr}[i]$ is $2$}
        {\eIf{$N\_temp$ is in $\textbf{NPS}$}
        {\emph{Pip2Enc\_Func(N\_temp)}}
        {\emph{Comb2Enc\_Func(N\_temp)}}}
        {\eIf{$N\_temp$ is in $\textbf{NPS}$}
        {\emph{Pip3Enc\_Func(N\_temp)}}
        {\emph{Comb3Enc\_Func(N\_temp)}}}}
    \eIf{$sys$ is $0$}
        {\emph{CombEnc\_Reg\_Func($N$)}}
        {\eIf{$pipln\_bndry$ is $0$} 
        {\emph{CombSysEnc\_Reg\_Func($N$)}} 
        {\emph{PipSysEnc\_Reg\_Func($N$)}}}
\textbf{Return} $T_c$
\caption{Execution flow of the proposed polar compiler}
\label{Alg:SW}
\end{algorithm}
\section{Implementation Results and Comparison}
\label{sec_result}
In this paper, all polar encoders are described using VHDL coding, while logic synthesis, technology mapping, and place-and-route are performed for Xilinx Artix-7 XC7A200T-2FBG676 FPGA. 
In order to enable comparison between various schemes, we define encoding latency as the time needed to encode a frame in both clock cycles (CCs) and seconds. The coded throughput of MK polar codes is calculated as $N.f$ bps. Similar to \cite{rezaei2023high, Coppolino, Rezaei2022MK}, we utilize the approach presented in \cite{bioglio} to determine the kernel ordering that yields the highest error-correction performance of MK codes as shown in Fig. \ref{fig:MKCompSCL2}.
\subsection{Time Efficiency of the Polar Compiler}
\begin{figure*}
    \centering
    \includegraphics[width=2.1\columnwidth]{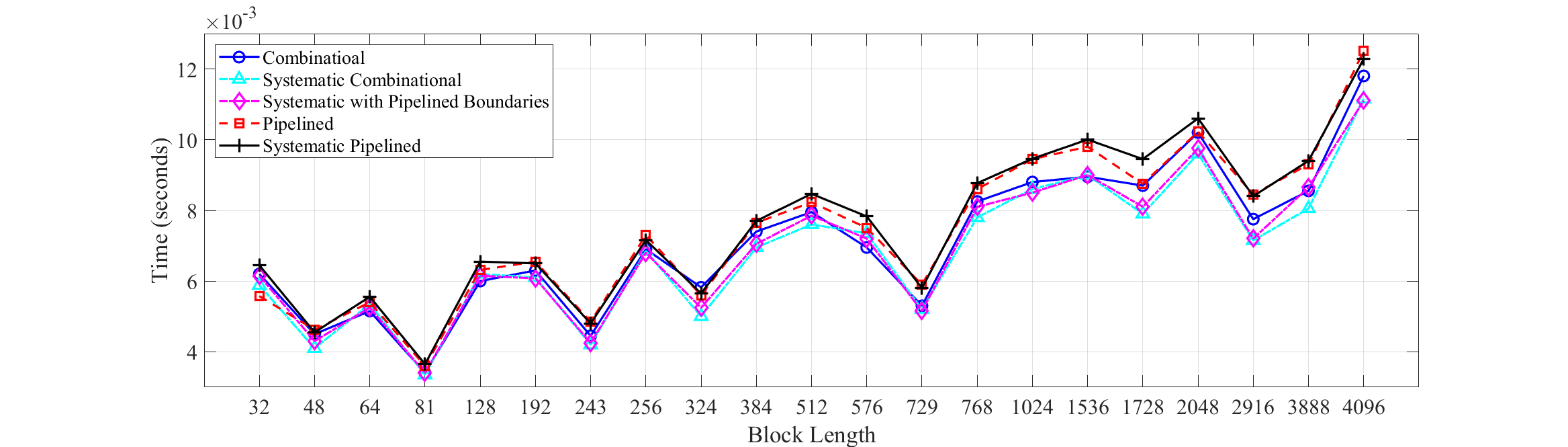}
    \caption{The CPU's elapsed time required for generating the hardware description of polar encoders of different sizes.}
    \label{fig:RunTime}
\end{figure*}

The time efficiency of the proposed polar compiler is assessed on an AMD Ryzen 7 PRO 5850U x64 CPU operating at a frequency of 1.90 GHz. The time required to generate all necessary VHDL files for polar encoders of varying sizes is illustrated in Fig. \ref{fig:RunTime}. Each data point is the average value obtained from running the proposed compiler $20$ times. The CPU compilation time is directly proportional to the number and complexity of the source codes, which is primarily influenced by the number of kernels in the kernel sequence. As illustrated in Fig. \ref{fig:RunTime}, the compilation time for Ar{\i}kan's encoders is greater than that of similarly sized MK encoders. This is attributed to the fact that Ar{\i}kan's encoders only comprise binary kernels, resulting in a longer kernel sequence.
For most polar codes with sizes up to $N=4096$, the compilation time remains under $12.5$ milliseconds, with a trend of nearly linear increase in response to the block length. Therefore, utilizing the proposed compiler represents an efficient approach to generating all necessary VHDL files for encoders with different architectures.
\subsection{Performance Evaluation and Comparison}
Table \ref{tab:MKEnc} presents the FPGA utilization and performance metrics of diverse non-systematic and systematic MK encoders, encompassing all conceivable kernel orderings, including pure-binary and pure-ternary codes. The utilization of look-up tables (LUTs) in any given code is closely tied to both the quantity and type of kernels that have been employed to build it. On the other hand, the number of occupied registers is influenced by the length of the code and the number of pipeline stages that have been utilized. As previously stated, the number of occupied registers is derived as $(\mathcal{P}+2)\cdot N$. In terms of encoding performance, it is noteworthy that the frequency of both non-systematic and systematic schemes exhibits a modest decline with an increase in code length. This is attributed to the larger number of kernels and interconnects involved in the process. However, the encoding performance experiences a significant increase proportional to the length of the code. By employing systematic encoders, the frequency and throughput are nearly halved as we use two consecutive non-systematic encoders. 
\label{sec_data}
\begin{table*}
\centering
\caption{Post-fitting results of various combinational non-systematic and systematic MK polar encoders}
\begin{tabular}{clccccccccccccc}
\hline
 &&&&\multicolumn{2}{c}{Non-systematic}&&&&&&\multicolumn{2}{c}{Systematic} \\\cline{3-8}  \cline{10-15} 
\begin{tabular}[c]{@{}c@{}}Block \\Length  \end{tabular}  &Kernel Order& LUTs  & Reg. & 
\begin{tabular}[c]{@{}c@{}}L \\(CCs)  \end{tabular}&
\begin{tabular}[c]{@{}c@{}}L \\($\mu$s)  \end{tabular}&
\begin{tabular}[c]{@{}c@{}}f \\(MHz)  \end{tabular}&
\begin{tabular}[c]{@{}c@{}}T/P \\(Gbps) \end{tabular}&&LUTs  & Reg. & 
\begin{tabular}[c]{@{}c@{}}L \\(CCs)  \end{tabular}&
\begin{tabular}[c]{@{}c@{}}L \\($\mu$s)  \end{tabular}&
\begin{tabular}[c]{@{}c@{}}f \\(MHz)  \end{tabular}&
\begin{tabular}[c]{@{}c@{}}T/P \\(Gbps) \end{tabular}\\\hline
192&\{3,2,2,2,2,2,2\}& 466 & 384 & 1 & 3.88 & 285 & 53.44&&906&384&1&5.85&171&32.06\\
256&\{2,2,2,2,2,2,2,2\}& 552 & 512 & 1 &3.7&270&67.5&&1115&512&1&5.68&176&44\\
243&\{3,3,3,3,3\}         & 641 &  486 & 1 &3.7&270&64.07&&1227&486&1&7.3&137&32.51\\
324&\{2,2,3,3,3,3\} &785&648&1&3.94&254&80.37&&1544& 648 & 1 &8.33&120& 37.9\\
384&\{3,2,2,2,2,2,2,2\}   &1007&768&1&3.65&274&102.75 &&1991&768&1&6.62&151&56.62\\
576&\{2,2,2,2,2,2,3,3\}   &1338&1152&1&4.33&231 &129.94&&3006&1152&1&7.94&126&70.87\\
1024&\{2,2,2,2,2,2,2,2,2,2\} & 2735 & 2048 &1&4.57& 219&219&&5369&2048&1&8.62&116&116\\
\hline
\end{tabular}
\label{tab:MKEnc}
\end{table*}

Table \ref{tab:MKEncPipOnestg} presents the post-fitting results of two systematic encoders after employing pipeline registers at the stage boundaries. Obviously, the frequencies and throughputs of this scheme bear a remarkable resemblance to those of their non-systematic counterparts. However, the marginal decrease in performance is attributed to the latency overhead of the pipeline registers.
\begin{table}
\centering
\caption{Post-fitting results of two systematic MK encoders after inserting pipeline registers at stage boundaries.}
\begin{tabular}{ccccccc}
\hline
\begin{tabular}[c]{@{}c@{}}Block \\Length  \end{tabular} & LUTs  & Reg. 
&
\begin{tabular}[c]{@{}c@{}}L \\(CCs)  \end{tabular}&
\begin{tabular}[c]{@{}c@{}}L \\(ns)  \end{tabular}&
\begin{tabular}[c]{@{}c@{}}f \\(MHz)  \end{tabular}&
\begin{tabular}[c]{@{}c@{}}T/P \\(Gbps) \end{tabular}\\\hline
324&1603&972&2&8.55&234&74.04\\
1024&5402&3072&2&9.43&212&212\\
\hline
\end{tabular}
\label{tab:MKEncPipOnestg}
\end{table}
The FPGA utilization and performance parameters of a polar encoder of size $N=1024$, as evaluated for various numbers of pipeline stages, are summarized in Table \ref{tab:MKEncPipvar}. 
\begin{table}
\centering
\caption{Post-fitting results of various pipelined MK polar encoders using $\mathcal{P}$ pipeline stages.}
\begin{tabular}{cccccccc}
\hline
\begin{tabular}[c]{@{}c@{}}Block \\Length  \end{tabular} &$\mathcal{P}$& LUTs  & Reg. 
&
\begin{tabular}[c]{@{}c@{}}L \\(CCs)  \end{tabular}&
\begin{tabular}[c]{@{}c@{}}L \\(ns)  \end{tabular}&
\begin{tabular}[c]{@{}c@{}}f \\(MHz)  \end{tabular}&
\begin{tabular}[c]{@{}c@{}}T/P \\(Gbps) \end{tabular}\\\hline
1024& 0 & 2735 & 2048 &1&4.57&219&219\\
1024& 1 & 2582 & 3072 &2&7.3&274&274\\
1024& 2 & 2051 & 4096 &3&10.87&276&276\\
1024& 4 & 2561 & 6144&5&17.73&282&282\\
1024& 9 & 5121 & 11264&10&35.34&283&283\\
\hline
\end{tabular}
\label{tab:MKEncPipvar}
\end{table}
With the exception of the deeply-pipelined scenario, the quantity of occupied LUTs falls within a similar range, while the number of registers scales proportionally to the number of pipeline stages. In the case of deeply-pipelined encoder, each XOR gate is sandwiched by two registers, resulting in a substantial increase in the number of LUTs. When considering frequency and throughput, it is notable that partially-pipelined encoders operate within a closely comparable frequency range and achieve comparable levels of throughput, thereby rendering the partially-pipelined schemes more appealing.

The results of implementing partially-pipelined non-systematic and systematic encoders, including pipelined boundary stages, are shown in Table \ref{tab:MKEncPip}. It is clear from the results that by utilizing roughly double resources, the systematic encoder achieves similar throughputs and operates at comparable frequencies with its non-systematic counterpart. The slight difference in performance is attributed to the delay caused by the additional pipeline stage inserted between two consecutive non-systematic encoders.
\begin{table*}
\centering
\caption{Post-fitting results of various partially-pipelined MK polar encoders}
\begin{tabular}{clccccccccccccc}
\hline
 &&&&\multicolumn{2}{c}{Non-systematic}&&&&&&\multicolumn{2}{c}{Systematic} \\\cline{3-8}  \cline{10-15} 
\begin{tabular}[c]{@{}c@{}}Block \\Length  \end{tabular}  &Kernel Order& LUTs  & Reg. & 
\begin{tabular}[c]{@{}c@{}}L \\(CCs)  \end{tabular}&
\begin{tabular}[c]{@{}c@{}}L \\(ns)  \end{tabular}&
\begin{tabular}[c]{@{}c@{}}f \\(MHz)  \end{tabular}&
\begin{tabular}[c]{@{}c@{}}T/P \\(Gbps) \end{tabular}&&LUTs  & Reg. & 
\begin{tabular}[c]{@{}c@{}}L \\(CCs)  \end{tabular}&
\begin{tabular}[c]{@{}c@{}}L \\($\mu$s)  \end{tabular}&
\begin{tabular}[c]{@{}c@{}}f \\(MHz)  \end{tabular}&
\begin{tabular}[c]{@{}c@{}}T/P \\(Gbps) \end{tabular}\\\hline
192&\{3,2,2,2,2,2,2\}     &417&960&4&14.39&278&52.12&&833&1728&8&29.4&272&51\\
243&\{3,3,3,3,3\}         &811&1458&5&17.8&281&66.68&&1621&2673&10&36.1&277&65.73\\
324&\{2,2,3,3,3,3\}      &1027&1944&5&18.05&277&87.64&&2053&3564&10&36.23&276&87.33\\
1024&\{2,2,2,2,2,2,2,2,2,2\} &2561&6144&5&17.73&282&282&&5121&11264&10&36.1&277&277\\\hline
\end{tabular}
\label{tab:MKEncPip}
\end{table*}
In Table \ref{tab:MKEncPipComp}, a comparison between the implementation results of a deeply-pipelined non-systematic encoder from \cite{Zhong2018} and the proposed non-systematic encoder with only one pipeline stage is shown. 
Notably, with the same amount of logic utilized, the proposed encoder operates at a $22.6\%$ lower frequency while achieving a $50.2\%$ higher throughput. It is worth mentioning that \cite{Zhong2018} does not include the input and output registers.
\begin{table}
\centering
\caption{Comparison of the FPGA implementation of the deeply-pipelined encoder of \cite{Zhong2018} to the proposed one-stage pipelined encoder.}
\begin{tabular}{cccccccccccc}
\hline Encoder&\begin{tabular}[c]{@{}c@{}}Block \\Length  \end{tabular} &\begin{tabular}[c]{@{}c@{}}Tech.\\(nm)  \end{tabular}&LUTs  & Reg.&
\begin{tabular}[c]{@{}c@{}}f \\(MHz)  \end{tabular}&
\begin{tabular}[c]{@{}c@{}}T/P \\(Gbps) \end{tabular}\\\hline
\cite{Zhong2018}&1024&28&2628&1025&356.223&182.38\\
This work&1024&28&2582&3072&274&274\\\hline
\end{tabular}
\label{tab:MKEncPipComp}
\end{table}

With all parameters scaled to $28$nm technology node, Table \ref{tab:MKEncPipComp2} presents a comparison between the proposed non-systematic combinational and partially-pipelined ($\mathcal{P}=5$) encoders and the fastest state-of-the-art ASIC encoders from \cite{SHIH2018292,Zhong2020,7991021}. We have granted the latter encoders an advantage, given that longer codes typically result in superior performance, and they are executed by ASIC, which offers greater performance than FPGA. Clearly, all encoders operate at frequencies that are one to two orders of magnitude higher while providing one to two orders of magnitude lower throughput.
The proposed pipelined encoder, on the other hand, achieves an impressive throughput of up to $1080$ Gbps. Additionally, all encoders of this paper support $83$ different block lengths, which is significantly more than the $15$ different codes supported by encoders of \cite{SHIH2018292,Zhong2020,7991021}.\begin{table}
\centering
\caption{Comparison of various non-systematic encoders.}
\begin{tabular}{ccccccc}
\hline Encoder&\begin{tabular}[c]{@{}c@{}}Block \\Length  \end{tabular}&\begin{tabular}[c]{@{}c@{}}IC \\Type  \end{tabular}
&\begin{tabular}[c]{@{}c@{}}Tech.\\(nm)  \end{tabular}&
\begin{tabular}[c]{@{}c@{}}f \\(MHz)  \end{tabular}&
\begin{tabular}[c]{@{}c@{}}T/P \\(Gbps) \end{tabular}&\begin{tabular}[c]{@{}c@{}}Supp.\\Codes  \end{tabular} \\ \hline
\cite{SHIH2018292}$^*$&4096&ASIC&65&2870&3.87&15\\
\cite{Zhong2020}$^*$&4096&ASIC&65&1160.7&37.43&15\\
\cite{7991021}$^*$&8192&ASIC&40&14285.7&228.57&15\\
\begin{tabular}[c]{@{}c@{}}This work\\(comb.)  \end{tabular} &4096&FPGA&28&162&648&83\\
\begin{tabular}[c]{@{}c@{}}This work\\(pipelined)  \end{tabular}&4096&FPGA&28&270&1080&83\\\hline
\end{tabular}
\flushleft 
\footnotesize{$~~~^*$ As in \cite{rezaei2023high}, the performance parameters are normalized to 28 nm CMOS technology using scaling techniques from \cite{GiardTechMap}.}
\label{tab:MKEncPipComp2}
\end{table}
\subsection{I/O Bounded Encoding}
The proposed set of architectures demands a substantial amount of throughput at the input and output of the encoder, particularly for deeply-pipelined architectures. Take, for example, a polar encoder with a size of $N=4096$, which necessitates a bus capable of delivering a staggering data rate of $2160$ Gbps for transferring the input and output data.
Thanks to cutting-edge Ultrascale technology \cite{Xilinx2014} and Altera Generation 10 \cite{Won2010MeetingTP}, such high data rates can be provided. For instance, a specific model of Xilinx Zynq UltraScale+ family features an impressive array of 96 GTX transceivers, with each capable of reaching a maximum theoretical data rate of 32.75 Gbps, resulting in an overall attainable data rate of over $3$ Tbps.
Given the need for high data rates in deeply-pipelined architectures, partially-pipelined architectures currently hold more appeal, especially when employing FPGA technology.
\section{Conclusion}
\label{sec_conc}
The article is presented a novel unrolled hardware architecture designed for high-throughput encoding of multi-kernel polar codes. The proposed approach is capable of encoding polar codes that are constructed with binary ($2\times 2$), ternary ($3\times 3$), or binary-ternary mixed kernels. The methodology provides an effective solution for achieving the desired balance between throughput and resource consumption, and offers remarkable flexibility in terms of code length. The proposed approach can generate five distinct architectures targeting systematic or non-systematic encoders. The FPGA post-fitting results for different block lengths and kernel orderings are reported, demonstrating the effectiveness of the proposed architectures. Notably, a partially-pipelined polar encoder of size $N=4096$ is shown to achieve a remarkable throughput of $1080$ Gbps.

Ultimately, we have developed a polar compiler in Python, which has the capability to automatically produce the necessary VHDL files essential for the FPGA implementation of our proposed encoders. The proposed compiler effortlessly generates all the required VHDL modules without any manual intervention.

\section*{Acknowledgment}
This research has been supported by the Academy of Finland, 6G Flagship program under Grant 346208. 


\begin{IEEEbiography}[{\includegraphics[width=1in,height=1.25in,clip,keepaspectratio]{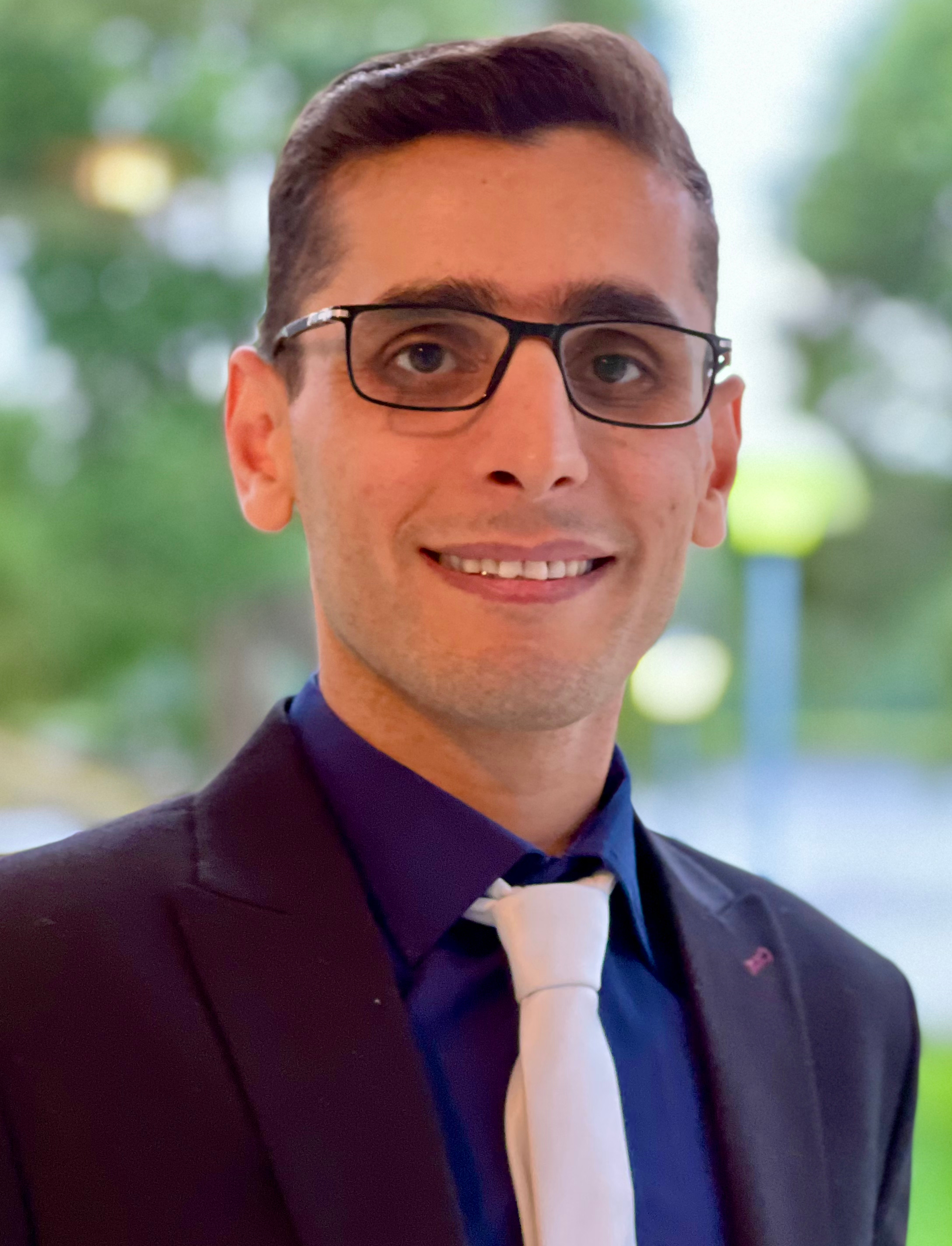}}]{Hossein Rezaei} (Graduate Student Member, IEEE) received his M.Sc degree in digital electronics from Iran University of Science and Technology, Tehran, Iran, in 2016. With over five years of experience as an FPGA/SoC designer in the industry, he has started his doctoral studies in Communications Engineering since 2020 at the University of Oulu, Oulu, Finland. He is also working as a senior SoC design engineer at Nokia, Oulu, Finland. His current research interests include design and implementation of error-correcting algorithms with a focus on polar codes, VLSI design for digital signal processing, semantic-based end-to-end transmission systems, and implementation of communication systems on embedded platform. 
\end{IEEEbiography}

\begin{IEEEbiography}[{\includegraphics[width=1in,height=1.25in,clip,keepaspectratio]{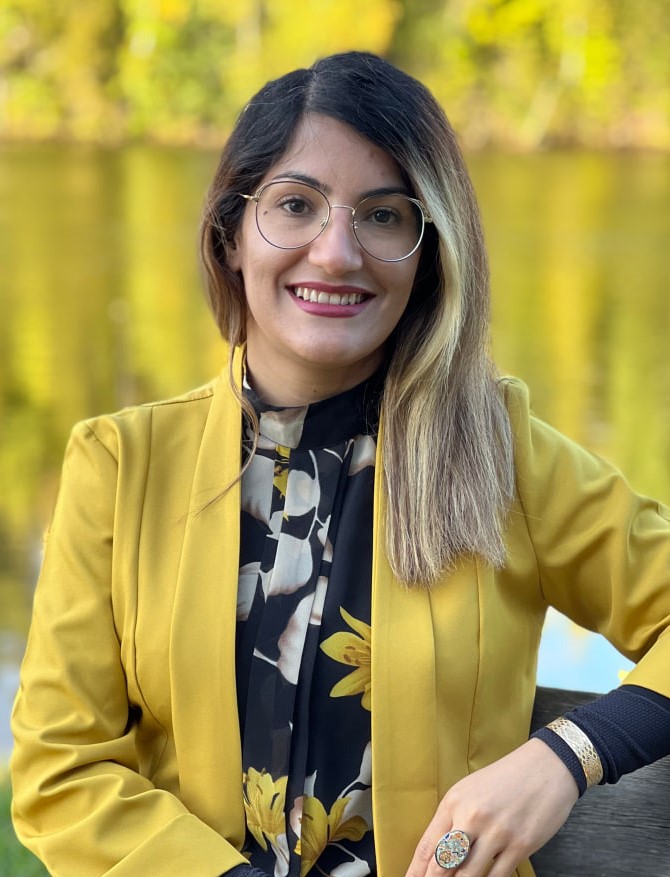}}]{Elham Abbasi} (Student Member, IEEE) received her M.Sc degree in electronics engineering from Semnan University, Semnan, Iran, in 2018. Following her graduation, she pursued a career as an FPGA engineer at Shahid Chamran University in Ahvaz, Iran, where she spent two years honing her skills and gaining valuable experience in the field. Her current research interests include design and implementation of error-correcting algorithms, VLSI design for digital signal processing, and implementation of communication systems on FPGA. Currently, she is a self-funded researcher based in Oulu, Finland.
\end{IEEEbiography}

\begin{IEEEbiography}[{\includegraphics[width=1in,height=1.25in,clip,keepaspectratio]{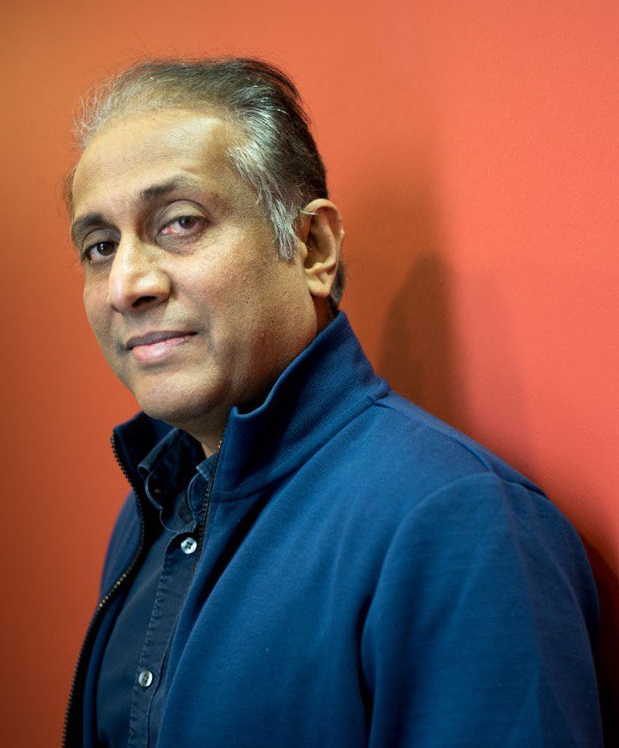}}]{Nandana Rajatheva}
(Senior Member, IEEE) received the B.Sc. (Hons.) degree in electronics and telecommunication engineering from the University of Moratuwa, Sri Lanka, in 1987, and the M.Sc. and Ph.D. degrees from the University of Manitoba, Winnipeg, MB, Canada, in 1991 and 1995, respectively. He is currently a Professor with the Centre for Wireless Communications, University of Oulu, Finland. During his graduate studies, he was a Canadian Commonwealth Scholar in Manitoba. From 1995 to 2010, he held a professor/associate professor positions with the University of Moratuwa and the Asian Institute of Technology, Thailand. He is currently leading the AI-driven Air Interface Design Task in Hexa-X EU Project. He has coauthored more than 200 referred articles published in journals and in conference proceedings. His research interests include physical layer in beyond $5$G, machine learning for PHY and MAC, integrated sensing and communications as well as channel coding. 
\end{IEEEbiography}

\begin{IEEEbiography}[{\includegraphics[width=1in,height=1.25in,clip,keepaspectratio]{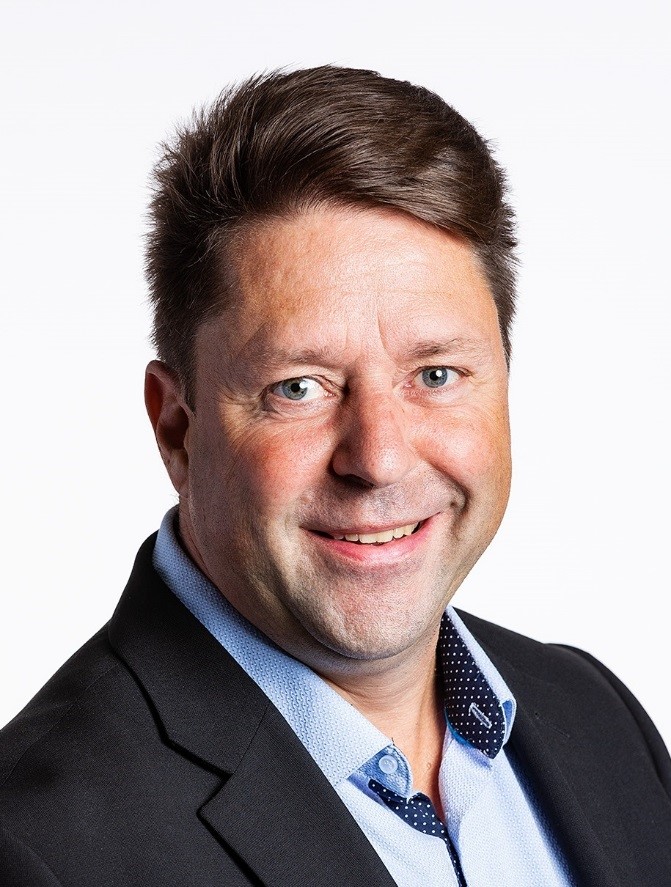}}]{Matti Latva-aho}
(Senior Member, IEEE) received the M.Sc., Lic.Tech. and Dr. Tech (Hons.) degrees in Electrical Engineering from the University of Oulu, Finland in 1992, 1996 and 1998, respectively. From 1992 to 1993, he was a Research Engineer at Nokia Mobile Phones, Oulu, Finland after which he joined Centre for Wireless Communications (CWC) at the University of Oulu. Prof. Latva-aho was Director of CWC during the years 1998-2006 and Head of Department for Communication Engineering until August 2014. Currently he serves as Academy of Finland Professor and is Director for National 6G Flagship Programme. He is also a Global Fellow with Tokyo University. His research interests are related to mobile broadband communication systems and currently his group focuses on 6G systems research. Prof. Latva-aho has published over 500 conference or journal papers in the field of wireless communications. He received Nokia Foundation Award in 2015 for his achievements in mobile communications research.
\end{IEEEbiography}

\begin{thebibliography}{10}
\providecommand{\url}[1]{#1}
\csname url@samestyle\endcsname
\providecommand{\newblock}{\relax}
\providecommand{\bibinfo}[2]{#2}
\providecommand{\BIBentrySTDinterwordspacing}{\spaceskip=0pt\relax}
\providecommand{\BIBentryALTinterwordstretchfactor}{4}
\providecommand{\BIBentryALTinterwordspacing}{\spaceskip=\fontdimen2\font plus
\BIBentryALTinterwordstretchfactor\fontdimen3\font minus
  \fontdimen4\font\relax}
\providecommand{\BIBforeignlanguage}[2]{{%
\expandafter\ifx\csname l@#1\endcsname\relax
\typeout{** WARNING: IEEEtran.bst: No hyphenation pattern has been}%
\typeout{** loaded for the language `#1'. Using the pattern for}%
\typeout{** the default language instead.}%
\else
\language=\csname l@#1\endcsname
\fi
#2}}
\providecommand{\BIBdecl}{\relax}
\BIBdecl

\bibitem{Arikan}
E.~Arikan, ``Channel polarization: A method for constructing capacity-achieving
  codes for symmetric binary-input memoryless channels,'' \emph{IEEE
  Transactions on information Theory}, vol.~55, no.~7, pp. 3051--3073, 2009.

\bibitem{tal2015list}
I.~Tal and A.~Vardy, ``List decoding of polar codes,'' \emph{IEEE Transactions
  on Information Theory}, vol.~61, no.~5, pp. 2213--2226, 2015.

\bibitem{niu2012crc}
K.~Niu and K.~Chen, ``Crc-aided decoding of polar codes,'' \emph{IEEE
  communications letters}, vol.~16, no.~10, pp. 1668--1671, 2012.

\bibitem{3GPP}
3rd Generation Partnership Project~(3GPP), \emph{5G; NR; Multiplexing and
  Channel Coding}.\hskip 1em plus 0.5em minus 0.4em\relax 3GPP document 38.212
  V.15.3.0, 2018.

\bibitem{7169326}
C.~Zhang, J.~Yang, X.~You, and S.~Xu, ``Pipelined implementations of polar
  encoder and feed-back part for sc polar decoder,'' in \emph{2015 IEEE
  International Symposium on Circuits and Systems (ISCAS)}, 2015, pp.
  3032--3035.

\bibitem{7482677}
G.~Sarkis, I.~Tal, P.~Giard, A.~Vardy, C.~Thibeault, and W.~J. Gross,
  ``Flexible and low-complexity encoding and decoding of systematic polar
  codes,'' \emph{IEEE Transactions on Communications}, vol.~64, no.~7, pp.
  2732--2745, 2016.

\bibitem{6951410}
H.~Yoo and I.-C. Park, ``Partially parallel encoder architecture for long polar
  codes,'' \emph{IEEE Transactions on Circuits and Systems II: Express Briefs},
  vol.~62, no.~3, pp. 306--310, 2015.

\bibitem{9063643}
W.~Song, Y.~Shen, L.~Li, K.~Niu, and C.~Zhang, ``A general construction and
  encoder implementation of polar codes,'' \emph{IEEE Transactions on Very
  Large Scale Integration (VLSI) Systems}, vol.~28, no.~7, pp. 1690--1702,
  2020.

\bibitem{SHIH2018292}
\BIBentryALTinterwordspacing
X.-Y. Shih, P.-C. Huang, and H.-R. Chou, ``Vlsi design and implementation of a
  reconfigurable hardware-friendly polar encoder architecture for emerging
  high-speed 5g system,'' \emph{Integration}, vol.~62, pp. 292--300, 2018.
  [Online]. Available:
  \url{https://www.sciencedirect.com/science/article/pii/S0167926017301232}
\BIBentrySTDinterwordspacing

\bibitem{7991021}
X.-Y. Shih and P.-C. Huang, ``Vlsi design of an ultra-high-speed polar encoder
  architecture using 16-parallel radix-2 processing engines for next-generation
  5g applications,'' in \emph{2017 IEEE International Conference on Consumer
  Electronics - Taiwan (ICCE-TW)}, 2017, pp. 113--114.

\bibitem{Zhong2020}
Z.~Zhong, W.~J. Gross, Z.~Zhang, X.~You, and C.~Zhang, ``Polar compiler:
  Auto-generator of hardware architectures for polar encoders,'' \emph{IEEE
  Transactions on Circuits and Systems I: Regular Papers}, vol.~67, no.~6, pp.
  2091--2102, 2020.

\bibitem{Coppolino}
G.~Coppolino, C.~Condo, G.~Masera, and W.~J. Gross, ``A multi-kernel multi-code
  polar decoder architecture,'' \emph{IEEE Transactions on Circuits and Systems
  I: Regular Papers}, vol.~65, no.~12, pp. 4413--4422, 2018.

\bibitem{Rezaei2022}
H.~Rezaei, V.~Ranasinghe, N.~Rajatheva, M.~Latva-aho, G.~Park, and O.-S. Park,
  ``Implementation of ultra-fast polar decoders,'' in \emph{2022 IEEE
  International Conference on Communications Workshops (ICC Workshops)}, 2022,
  pp. 235--241.

\bibitem{rezaei2022combinational}
H.~Rezaei, N.~Rajatheva, and M.~Latva-aho, ``A combinational multi-kernel
  decoder for polar codes,'' \emph{arXiv preprint arXiv:2211.08778}, 2022.

\bibitem{Rezaei2022MK}
------, ``Low-latency multi-kernel polar decoders,'' \emph{IEEE Access},
  vol.~10, pp. 119\,460--119\,474, 2022.

\bibitem{rezaei2023high}
------, ``High-throughput rate-flexible combinational decoders for multi-kernel
  polar codes,'' \emph{arXiv preprint arXiv:2301.10445}, 2023.

\bibitem{mori2009performance}
R.~Mori and T.~Tanaka, ``Performance of polar codes with the construction using
  density evolution,'' \emph{IEEE Communications Letters}, vol.~13, no.~7, pp.
  519--521, 2009.

\bibitem{Shin2012}
K.-W. Shin and H.~ju~Kim, ``A multi-mode ldpc decoder for ieee 802.16e mobile
  wimax,'' \emph{Journal of Semiconductor Technology and Science}, vol.~12, pp.
  24--33, 2012.

\bibitem{Zhang2012}
L.~Zhang, Z.~Zhang, and X.~Wang, ``Polar code with block-length n = 3n,'' in
  \emph{2012 International Conference on Wireless Communications and Signal
  Processing (WCSP)}, 2012, pp. 1--6.

\bibitem{8746303}
L.~Cheng, W.~Zhou, and L.~Zhang, ``Hybrid multi-kernel construction of polar
  codes,'' in \emph{2019 IEEE 89th Vehicular Technology Conference
  (VTC2019-Spring)}, 2019, pp. 1--5.

\bibitem{arikan2011systematic}
E.~Arikan, ``Systematic polar coding,'' \emph{IEEE communications letters},
  vol.~15, no.~8, pp. 860--862, 2011.

\bibitem{nane2015survey}
R.~Nane, V.-M. Sima, C.~Pilato, J.~Choi, B.~Fort, A.~Canis, Y.~T. Chen,
  H.~Hsiao, S.~Brown, F.~Ferrandi \emph{et~al.}, ``A survey and evaluation of
  fpga high-level synthesis tools,'' \emph{IEEE Transactions on Computer-Aided
  Design of Integrated Circuits and Systems}, vol.~35, no.~10, pp. 1591--1604,
  2015.

\bibitem{PythonCode}
H.~Rezaei, E.~Abbasi, and N.~Rajatheva, \emph{Polar Encoder Compiler}.\hskip
  1em plus 0.5em minus 0.4em\relax Accessed, May 2023. [Online]. Available:
  https://github.com/hosseinrezaeii91/Polar-Encoder-Compiler, 2023.

\bibitem{bioglio}
V.~Bioglio, F.~Gabry, I.~Land, and J.-C. Belfiore, ``Minimum-distance based
  construction of multi-kernel polar codes,'' in \emph{GLOBECOM 2017-2017 IEEE
  Global Communications Conference}.\hskip 1em plus 0.5em minus 0.4em\relax
  IEEE, 2017, pp. 1--6.

\bibitem{gabry2017}
F.~Gabry, V.~Bioglio, I.~Land, and J.-C. Belfiore, ``Multi-kernel construction
  of polar codes,'' in \emph{2017 IEEE International Conference on
  Communications Workshops (ICC Workshops)}.\hskip 1em plus 0.5em minus
  0.4em\relax IEEE, 2017, pp. 761--765.

\bibitem{Zhong2018}
Z.~Zhong, X.~You, and C.~Zhang, ``Auto-generation of pipelined hardware designs
  for polar encoder,'' in \emph{2018 China Semiconductor Technology
  International Conference (CSTIC)}, 2018, pp. 1--4.

\bibitem{GiardTechMap}
P.~Giard, A.~Balatsoukas-Stimming, T.~C. Müller, A.~Bonetti, C.~Thibeault,
  W.~J. Gross, P.~Flatresse, and A.~Burg, ``Polarbear: A 28-nm fd-soi asic for
  decoding of polar codes,'' \emph{IEEE Journal on Emerging and Selected Topics
  in Circuits and Systems}, vol.~7, no.~4, pp. 616--629, 2017.

\bibitem{Xilinx2014}
Xilinx, \emph{"UltraScale architecture and product overview,"}.\hskip 1em plus
  0.5em minus 0.4em\relax Product Specification, Dec. 2014.

\bibitem{Won2010MeetingTP}
Altera, ``Meeting the performance and power imperative of the zettabyte era
  with generation 10.''\hskip 1em plus 0.5em minus 0.4em\relax White Paper,
  2013.

\end{thebibliography}
\end{document}